\documentclass[aps, prb, reprint, superscriptaddress, floatfix]{revtex4-2}

\usepackage{xcolor}
\usepackage{booktabs}
\usepackage{graphicx}% Include figure files
\usepackage{dcolumn}% Align table columns on decimal point
\usepackage{amssymb}
\usepackage{amsmath}
\usepackage{hyperref}% add hypertext capabilities
\hypersetup{colorlinks=true, citecolor=blue, linkcolor=blue, urlcolor=blue}
\usepackage[Symbolsmallscale]{upgreek}

\newcommand{\emc}{e$\upmu$c}

\newcolumntype{d}[1]{D{.}{.}{#1}}

\begin{document}

\title{Quantifying errors of the electron-proton/muon correlation functionals through the Kohn-Sham inversion of a two-component model system}

\author{Nahid Sadat Riyahi}
\affiliation{Department of Physical and Computational Chemistry, Shahid Beheshti University, Evin, Tehran 19839-69411, Iran}
\author{Mohammad Goli}
\email{m{\_}goli@ipm.ir}
\affiliation{School of Nano Science, Institute for Research in Fundamental Sciences (IPM), Tehran 19395-5531, Iran}
\author{Shant Shahbazian}
\email{sh{\_}shahbazian@sbu.ac.ir, sh-shahbazian@sharif.edu}
\affiliation{Department of Physics, Shahid Beheshti University, Evin, Tehran 19839-69411, Iran}
\affiliation{Department of Chemistry, Sharif University of Technology, Tehran 11155-9516, Iran}

\date{\today}

\begin{abstract}
The multi-component density functional theory is faced with the challenge of capturing various types of inter- and intra-particle exchange-correlation effects beyond those introduced by the conventional electronic exchange-correlation functionals. Herein, we focus on evaluating the electron-proton/muon correlation functionals appearing in molecular/condensed-phase systems where a proton/muon is treated as a quantum particle on equal footing with electrons, beyond the Born-Oppenheimer paradigm. Five recently developed local correlation functionals, i.e., the epc series and \emc-1, are selected and their performances are analyzed by employing a two-particle model that includes an electron and a positively charged particle (PCP) with a variable mass, interacting through Coulombic forces, within a double harmonic trap. Using the Kohn-Sham (KS) inversion procedure, the exact two-component KS characterization of the model is deduced and its properties are compared to those derived from the considered functionals. The analysis demonstrates that these local functionals achieve their original parameterization objectives to reproduce the one-PCP densities and the electron-PCP correlation energies, but all fall short of reproducing the underlying PCP correlation potentials correctly. Moreover, a comprehensive error analysis reveals that the density-driven errors have a non-negligible contribution to the success of the considered functionals. Overall, the study shows the strengths as well as shortcomings of the considered functionals hopefully paving the way for designing more robust functionals in the future.   
\end{abstract}


\maketitle

\section{Introduction}%
The last two decades have witnessed a new age in \textit{ab initio} computational study of the multi-component Coulombic quantum systems which has been recently called the multi-component quantum chemistry (MCQC) \cite{pavosevic_MulticomponentQuantumChemistry_2020}. The MCQC aims to extend the applicability domain of \textit{ab initio} QC to molecular systems and phenomena where not only electrons but also other particles are treated as quantum particles. Apart from the usual light nuclei like proton and its heavier isotopes, this includes elementary particles that are attached to molecular systems like positron or the positively charged muon (hereafter called muon for brevity) \cite{ache_ChemistryPositronPositronium_1972a,puska_TheoryPositronsSolids_1994,jean_PrinciplesApplicationsPositron_2003,bailey_PositronEmissionTomography_2005,gribakin_PositronmoleculeInteractionsResonant_2010,tuomisto_DefectIdentificationSemiconductors_2013,emami-razavi_ReviewExperimentalTheoretical_2021,patterson_MuoniumStatesSemiconductors_1988,storchak_QuantumDiffusionMuons_1998,blundell_SpinpolarizedMuonsCondensed_1999,nagamine_IntroductoryMuonScience_2003,blundell_MuonSpinRotationStudies_2004,clayden_MuonsChemistry_2013,hillier_MuonSpinSpectroscopy_2022}. While there is a “prehistory” for this extension \cite{thomas_ProtonicStructureMethane_1969,thomas_ProtonicStructureMolecules_1969,thomas_SelectionRulesProtonic_1970,pettitt_HartreeFockTheoryProton_1986,pettitt_SelfconsistentFieldProton_1987}, in one sense the renaissance of the MCQC started from the pioneering work of Tachikawa, Nakai, Shigeta and coworkers in 1998 \cite{tachikawa_ExtensionInitioMolecular_1998,tachikawa_FullVariationalMolecular_1998,shigeta_NonadiabaticMolecularTheory_1998,shigeta_DensityFunctionalTheory_1998}. These studies laid the stepping stone to attribute and optimize spin-spatial orbitals simultaneously to electrons as well as nuclei and/or other elementary particles in molecular systems. Since then various research groups have tried to bring the traditional hierarchical structure of the electronic \textit{ab initio} methodologies, i.e., first Hartree-Fock (HF) and then step by step more sophisticated post-HF methods, to the realm of the MCQC \cite{nakai_NuclearOrbitalMolecular_2007,gonzalez_TheoreticalInvestigationIsotope_2008,ishimoto_ReviewMulticomponentMolecular_2009,flores-moreno_LOWDINAnyParticle_2014,reyes_AnyParticleMolecular_2019,rodas_AnyParticleMolecular_2019,hammes-schiffer_NuclearElectronicOrbital_2021}. At first glance this may seem to be a computationally cumbersome but theoretically straightforward research program without any need for innovative theoretical elements. However, this is deceptive since new types of correlations emerge in such systems apart from the well-known electron-electron correlation \cite{helgaker_MolecularElectronicStructureTheory_2000,fulde_CorrelatedElectronsQuantum_2012}. 

Let us imagine a two-component system composed of $N$ electrons and a single positively charged particle (PCP), all within an external field. The MCHF wavefunction, as the starting point for \textit{ab initio} calculations, is a product of the electronic $N \times N$ Slater determinant and the spin-orbital attributed to the PCP, which is an uncorrelated description of electrons and the PCP. Neglecting all types of correlations among the involved particles leads to a significant difference in the observable results computed at the MCHF level compared to the exact solution of the MC-Schr\"{o}dinger equation \cite{cassam-chenai_DecouplingElectronsNuclei_2015,cassam-chenai_QuantumChemicalDefinition_2017a}. Inevitably, in subsequent steps in any conceived hierarchical structure of the MCQM, one must deal with both electron-electron and electron-PCP correlations and try to incorporate them efficiently into the wavefunction. Nevertheless, the electron-PCP correlation is not only quantitatively but also qualitatively different from the electron-electron correlation because an electron and a PCP form a distinguishable and \textit{attractively} interacting pair of particles with no operative exchange phenomenon, unlike electrons. In fact, two decades of experience reveal that the orbital-based \textit{ab initio} correlated methods, which are capable of recovering electron-electron correlation, are not suitable to recover the electron-PCP correlation efficiently and new methodological developments are necessary \cite{webb_MulticonfigurationalNuclearelectronicOrbital_2002,swalina_ExplicitDynamicalElectron_2006,chakraborty_InclusionExplicitElectronproton_2008,adamson_ModelingPositronsMolecular_2008,chakraborty_DensityMatrixFormulation_2008,pak_CalculationPositronAnnihilation_2009,ko_AlternativeWavefunctionAnsatz_2011,hoshino_RigorousNonBornOppenheimerTheory_2011,nishizawa_DevelopmentExplicitlyCorrelated_2012,nishizawa_EvaluationElectronRepulsion_2012,swalina_AnalysisElectronpositronWavefunctions_2012,sirjoosingh_ReducedExplicitlyCorrelated_2013,sirjoosingh_ReducedExplicitlyCorrelated_2013a,diaz-tinoco_GeneralizedAnyparticlePropagator_2013,suzuki_ElectronicSchrOdinger_2014,sirjoosingh_QuantumTreatmentProtons_2015,brorsen_NuclearelectronicOrbitalReduced_2015,ellis_DevelopmentMulticomponentCoupledCluster_2016,tsukamoto_DivideandconquerSecondorderProton_2016,ellis_InvestigationManyBodyCorrelation_2017,pavosevic_MulticomponentCoupledCluster_2019,pavosevic_MulticomponentCoupledCluster_2019a,pavosevic_MulticomponentEquationofmotionCoupled_2019,pavosevic_MulticomponentOrbitalOptimizedPerturbation_2020,fajen_SeparationElectronElectron_2020,muolo_NuclearelectronicAllparticleDensity_2020,harkonen_ManybodyGreenFunction_2020,brorsen_QuantifyingMultireferenceCharacter_2020,agostini_ExactFactorizationElectron_2020a,fajen_MulticomponentCASSCFRevisited_2021,chen_NucleusElectronCorrelation_2021}. This is also the case for the extension of density functional theory (DFT) to the MC systems as one of the most computationally cost-effective methodologies to deal with the many-body problem \cite{engel_DensityFunctionalTheory_2011}. While the theoretical foundations of the MCDFT were laid decades ago \cite{sander_SurfaceStructureElectronHole_1973a,sander_SurfaceStructureElectronHole_1973,kalia_SurfaceStructureElectronhole_1978,kryachko_FormulationVrepresentableDensity_1991,capitani_NonBornOppenheimer_1982,nalewajski_DensityFunctionalTheory_1982,nalewajski_InternalDensityFunctional_1984,nieminen_TwocomponentDensityfunctionalTheory_1985,boronski_ElectronpositronDensityfunctionalTheory_1986,li_TimedependentDensityfunctionalTheory_1986}, and refined since then \cite{gidopoulos_KohnShamEquationsMulticomponent_1998,barnea_DensityFunctionalTheory_2007,engel_IntrinsicdensityFunctionals_2007,chakraborty_PropertiesExactUniversal_2009,messud_DensityFunctionalTheory_2009,messud_GeneralizationInternalDensityfunctional_2011,gidopoulos_ElectronicNonadiabaticStates_2014,culpitt_MulticomponentDensityFunctional_2016,requist_ExactFactorizationBasedDensity_2016,suzuki_TimeDependentMulticomponentDensity_2018,requist_ExactFactorizationbasedDensity_2019,xu_FullquantumDescriptionsMolecular_2020}, properly designed electron-PCP correlation functionals have appeared only recently \cite{puska_ElectronpositronCarParrinelloMethods_1995,kreibich_MulticomponentDensityFunctionalTheory_2001,ito_FormulationNumericalApproach_2004,udagawa_IsotopeEffectPorphine_2006,imamura_ColleSalvettitypeCorrectionElectron_2008,chakraborty_DevelopmentElectronProtonDensity_2008,kreibich_MulticomponentDensityfunctionalTheory_2008,imamura_ExtensionDensityFunctional_2009,sirjoosingh_DerivationElectronProton_2011,sirjoosingh_MulticomponentDensityFunctional_2012,kuriplach_ImprovedGeneralizedGradient_2014,udagawa_ElectronnucleusCorrelationFunctional_2014,zubiaga_FullcorrelationSingleparticlePositron_2014,wiktor_TwocomponentDensityFunctional_2015,yang_DevelopmentPracticalMulticomponent_2017,brorsen_AlternativeFormsTransferability_2018,kolesov_DensityFunctionalTheory_2018,tao_MulticomponentDensityFunctional_2019,goli_TwocomponentDensityFunctional_2022}. Although it is not discussed in the present study, let us just briefly stress that another type of correlation, namely the PCP-PCP correlation, may also emerge if, for example, one considers quantum systems composed of $N$ electrons and $M$ PCPs when $M>1$. Probably, the most interesting example is hydrogen under extreme pressures and its infamous metallic phase \cite{mcmahon_PropertiesHydrogenHelium_2012,nellis_WignerHuntingtonLong_2013,gregoryanz_EverythingYouAlways_2020,silvera_PhasesHydrogenIsotopes_2021}, which is basically a strongly interacting system of electrons and protons (or deuteriums) \cite{xu_DensityfunctionalTheoryPair_1998,kitamura_QuantumDistributionProtons_2000,xu_DensityFunctionalTheory_2002,goncharov_ProtonDelocalizationExtreme_2007,pierleoni_TrialWaveFunctions_2008,mcmahon_GroundStateStructuresAtomic_2011,mcminis_MolecularAtomicPhase_2015,liao_ComparativeStudyUsing_2019,tenney_PossibilityMetastableAtomic_2020,gorelov_EnergyGapClosure_2020,monacelli_BlackMetalHydrogen_2021,niu_StableSolidMolecular_2023}. Accordingly, in an \textit{ab initio} study of an MC many-body quantum system, composed of quantum particles with different charges and masses, the diversity of various conceived correlations is enormous. Currently, in contrast to all attempts \cite{allen_VariationalMethodGround_1968,broyles_MethodsComputingDistribution_1969,barker_EffectivePotentialsComponents_1971,pokrant_NonzeroTemperatureVariational_1974,chakraborty_VariationalApproachGround_1982,dharma-wardana_DensityfunctionalTheoryHydrogen_1982,pietilainen_VariationalApproachTwocomponent_1984,bishop_TwocomponentFermiSystems_1987,lahoz_TwocomponentFermiSystems_1988,perrot_EquationStateTransport_1995,hetenyi_ApproximateSolutionVariational_2010,dharma-wardana_ElectronionIonionPotentials_2012}, there is no single unified scheme to treat all these correlations efficiently and much remains to be done in this area.       

In the meantime, the most computationally tractable \textit{ab initio} methodology to deal with medium to large MC molecular systems is the MCDFT and the solution of the MC Kohn-Sham (MCKS) equations. The recent success in introducing relatively accurate local and semi-local electron-positron \cite{puska_ElectronpositronCarParrinelloMethods_1995,kuriplach_ImprovedGeneralizedGradient_2014,zubiaga_FullcorrelationSingleparticlePositron_2014}, electron-proton \cite{yang_DevelopmentPracticalMulticomponent_2017,brorsen_AlternativeFormsTransferability_2018,tao_MulticomponentDensityFunctional_2019}, and electron-muon \cite{goli_TwocomponentDensityFunctional_2022}, correlation functionals are indeed promising. However, designing novel correlation functionals is by no means a straightforward task since the goal of building a simplified but reliable many-body model to start considering the electron-PCP correlation, similar to the homogeneous electron gas model used in the electronic DFT (eDFT) \cite{parr_DensityFunctionalTheoryAtoms_1994,mahan_ManyParticlePhysics_2000}, is not achieved yet. The only exception is the “delocalized” states of positrons in solids \cite{puska_TheoryPositronsSolids_1994}, for which the many-body homogeneous electron-positron gas model was developed long ago \cite{brinkman_ElectronHoleLiquidsSemiconductors_1973,arponen_ElectronLiquidCollective_1979,pietilainen_HypernettedchainTheoryCharged_1983,keldysh_ElectronholeLiquidSemiconductors_1986,lantto_VariationalTheoryMulticomponent_1987,drummond_QuantumMonteCarlo_2011}. This model has been employed to deduce local density approximations for the electron-positron correlation functional \cite{manninen_ElectronsPositronsMetal_1975,gupta_ElectronPositronDensities_1977,puska_DefectSpectroscopyPositrons_1983}. While the model has been extended to the case of a PCP with an arbitrary mass \cite{almbladh_ScreeningProtonElectron_1976,kallio_HypernettedChainTheory_1982,gondzik_ScreeningPositiveParticles_1985,stachowiak_ScreeningProtonMoving_1984,stachowiak_ScreeningProtonMoving_1987a,stachowiak_ScreeningProtonMoving_1987,deng_TwocomponentDensityFunctional_2023}, the “localized” nature of the heavy PCPs, e.g., proton and muon, in many MC systems cast some doubt on the model’s validity in such cases. As an alternative, the recently designed electron-proton correlation functionals have been largely based on extending the Colle-Salvetti approximation for the electron-electron correlation energy \cite{colle_ApproximateCalculationCorrelation_1975a,handy_ImportanceColleSalvetti_2009}, to the electron-proton correlation energy \cite{yang_DevelopmentPracticalMulticomponent_2017,brorsen_AlternativeFormsTransferability_2018,tao_MulticomponentDensityFunctional_2019}. Although some theoretical clarifications yet needed to be done in this procedure, the computational success of the series of designed functionals, called epc-17 \cite{yang_DevelopmentPracticalMulticomponent_2017}, epc-18 \cite{brorsen_AlternativeFormsTransferability_2018} and epc-19 \cite{tao_MulticomponentDensityFunctional_2019}, encourages this line of reasoning \cite{pavosevic_MulticomponentQuantumChemistry_2020,brorsen_MulticomponentDensityFunctional_2017,yang_MulticomponentTimeDependentDensity_2018,tao_DirectDynamicsNuclear_2021}. The \emc-1 functional \cite{goli_TwocomponentDensityFunctional_2022}, which is recently proposed to account for the electron-muon correlation effects, resembles the epc series mathematically, but simply designed based on a semi-empirical approach from the outset (see Appendix \ref{sec:a3}). Accordingly, it is desirable to check the intrinsic accuracy of these functional and try to uncover the reasons behind their success.  By the way, this is not a straightforward task, since in an \textit{ab initio} MCDFT calculation the outcomes, e.g., total energy or geometrical parameters, depend both on the qualities of the electronic exchange-correlation and the electron-PCP correlation functionals. A handful of studies on the possible inter-dependence of these two types of correlations have come to opposing views, and further studies are needed for more clarification \cite{sirjoosingh_MulticomponentDensityFunctional_2012,udagawa_ElectronnucleusCorrelationFunctional_2014,brorsen_AlternativeFormsTransferability_2018}. Nevertheless, all the mentioned electron-PCP functionals have been parameterized in \textit{ab initio} procedures on real molecules employing the same electronic exchange-correlation functionals that are used in the “single-component” eDFT without any re-parameterization. A reasonable doubt emerges as to what extent the electron-PCP correlation functionals “absorb” the errors inherent in the original design of the electronic exchange-correlation functionals. In fact, the error compensation of two seemingly independent functionals that are used in conjunction with \textit{ab initio} calculations is well-documented and has been utilized in the joint parameterization of the electronic exchange and correlation functionals \cite{kohanoff_ElectronicStructureCalculations_2006}. 

Currently, the “gold standard” for the quality of an electron-PCP correlation functional is its ability to overcome the \textit{overlocalization} of the uncorrelated one-PCP density (\textit{vide infra}). These overlocalized uncorrelated densities are derived from MCKS (and MCHF) wavefunctions in the absence of the electron-PCP correlation functional \cite{yang_DevelopmentPracticalMulticomponent_2017,brorsen_AlternativeFormsTransferability_2018,tao_MulticomponentDensityFunctional_2019,goli_TwocomponentDensityFunctional_2022}. However, these studies demonstrate that the “cure” of the overlocalization does not in itself guarantee the success of an electron-PCP correlation functional in reproducing the correct energetics. Thus, it is desirable to find new ways of gauging the “inherent” accuracy of the currently used electron-PCP correlation functionals. To reach this goal, it is preferable to apply the MCKS equations with the desired electron-PCP correlation functional to a model system that lacks the electron-electron and PCP-PCP correlations. Such a system is inevitably composed of just one electron and a PCP, all placed within an external field, to eliminate the unbound center of mass motion preventing concomitant complications \cite{bochevarov_ElectronNuclearOrbitals_2004}. 

The following section presents the basic theory of this model system and shows that the overlocalization problem persists in this system regardless of whether the PCP is a proton or a muon. Thus this simple model is a proper testing ground to evaluate the quality of the electron-PCP correlation functionals. Also, various energetic comparisons are done comparing the results of the exact MCKS with those derived from various local correlation functionals including, the electron-PCP correlation energy and the KS potentials of the electron and the PCP. Overall, the present study tries to evaluate the inherent qualities of the local electron-proton and electron-muon correlation functionals currently in use. 

\section{The model two-particle system}
\subsection{The theoretical basics}

The idea of using simple models to evaluate the correlation energies in many-body systems, i.e., the difference between the exact and the mean-field energies, is not new and its usefulness has been demonstrated many times previously \cite{moshinsky_HarmonicOscillatorModern_1996,sutherland_BeautifulModels70_2004,maruhn_SimpleModelsManyFermion_2014,march_ExactlySolvableModels_2016}. In the case of electron-electron correlation, probably the best-known and the most studied model is the two-electron harmonium atom, sometimes also called the Hooke’s atom, which is composed of two electrons interacting within a harmonic trap \cite{piela_IdeasQuantumChemistry_2007}. The model was first proposed by Kestner and Sinanoglu in 1962 \cite{kestner_StudyElectronCorrelation_1962}, as a simplified version of the real helium atom, and since then it has been studied by various researchers \cite{santos_CalculoAproximadoEnergia_1968,tuan_DoublePerturbationTheory_1969,white_PerturbationTheoryHooke_1970,benson_PerturbationEnergiesHooke_1970,kais_DimensionalScalingSymmetry_1989,samanta_CorrelationExactlySolvable_1990,ghosh_StudyCorrelationEffects_1991,taut_TwoElectronsExternal_1993,turbiner_TwoElectronsExternal_1994,king_ElectronCorrelationCusp_1996,zhu_SizeShapeEffects_1997,lamouche_TwoInteractingElectrons_1998,cioslowski_GroundStateHarmonium_2000,henderson_ElectronCorrelationArtificial_2001,romera_ElectronpairUncertaintyRelationships_2002,cyrnek_EnergySpectrumHarmonium_2003,oneill_WaveFunctionsTwoelectron_2003,mandal_TwoElectronsHarmonic_2003,amovilli_ApproximateAnsatzExpansion_2003,karwowski_Harmonium_2004,gill_ElectronCorrelationHooke_2005,katriel_EffectOnebodyPotential_2005,akbari_MomentumDensitySpatial_2007,ragot_CommentsHartreeFock_2008,loos_CorrelationEnergyTwo_2009,matito_PropertiesHarmoniumAtoms_2010,ebrahimi-fard_HarmoniumLaboratoryMathematical_2012,karwowski_BiconfluentHeunEquation_2014,nagy_InformationtheoreticAspectsFriction_2015,cioslowski_HarmoniumAtomsWeak_2017,galiautdinov_GroundStateExciton_2018,cioslowski_NaturalOrbitalsGround_2018a,rusin_PauliExclusionOperator_2021}. In contrast to the helium atom, the partial electronic Schr\"{o}dinger differential equation of the harmonium atom is separable into two ordinary differential equations \cite{kestner_StudyElectronCorrelation_1962}. One of them is equivalent to the Schr\"{o}dinger equation of the harmonic oscillator problem, and is analytically solvable, while the other belongs to the class of quasi-exactly solvable models \cite{kestner_StudyElectronCorrelation_1962,santos_CalculoAproximadoEnergia_1968,kais_DimensionalScalingSymmetry_1989,taut_TwoElectronsExternal_1993,ushveridze_QuasiExactlySolvableModels_1994}. Since the exact analytical and/or numerical solutions of these equations are available, the model is an ideal “laboratory” to study the electron correlation effects. One of the most interesting applications of this laboratory is the evaluation of the reliability of various proposed approximate electronic exchange-correlation functionals used in the eDFT. The exact non-interacting KS system of the harmonium atom and its components, e.g., the KS correlation energy and the exchange-correlation potential, are derivable from an inversion process. Hence, the exact components are comparable with {their approximate counterparts}, derived from the approximate exchange-correlation functionals \cite{burke_SemilocalDensityFunctionals_1995}. The pioneering studies of Laufer and Krieger in 1986 \cite{laufer_TestDensityfunctionalApproximations_1986}, as well as other researchers \cite{hall_ComparisonPathIntegral_1989,samanta_DensityfunctionalApproachCalculation_1991,samanta_StudyCorrelationKohn_1991,kais_DensityFunctionalsDimensional_1993,filippi_ComparisonExactApproximate_1994}, set the stage for subsequent studies on various aspects of the eDFT of the harmonium atom \cite{huang_LocalCorrelationEnergies_1997,taut_TwoElectronsExternal_1998,qian_PhysicsTransformationSchr_1998,march_DifferentialEquationGroundstate_1998,lam_VirialExchangeCorrelation_1998,hessler_SeveralTheoremsTimeDependent_1999,hessler_ErratumSeveralTheorems_1999,ivanov_ExactHighdensityLimit_1999,march_WavefunctionWhenAntiparallel_2000,frydel_AdiabaticConnectionAccurate_2000,ludena_FunctionalNrepresentabilityDensity_2000,artemyev_DFTVariationalApproach_2002a,amovilli_ExactDensityMatrix_2003,holas_WaveFunctionsLoworder_2003,march_KineticEnergyDensity_2003,march_EffectiveOnebodyPotential_2004,ludena_ExactAnalyticTotal_2004,katriel_StudyAdiabaticConnection_2004,capuzzi_DifferentialEquationGroundstate_2005,ragot_ExactKohnShamHartreeFock_2006,gomez_ApplicationExactAnalytic_2006,zhu_ExactDensityFunctionals_2006,seidl_StrictlyCorrelatedElectrons_2007,katriel_NonlocalWignerlikeCorrelation_2007,coe_EntanglementDensityfunctionalTheory_2008,coe_ErratumEntanglementDensityfunctional_2010,sun_ExtensionNegativeValues_2009,gori-giorgi_StudyDiscontinuityExchangecorrelation_2009,seidl_AdiabaticConnectionNegative_2010,cioslowski_RobustValidationApproximate_2015,chauhan_StudyAdiabaticConnection_2017a,kooi_LocalGlobalInterpolations_2018,singh_SemianalyticalWavefunctionsKohn_2020}. Accordingly, it is desirable to have a similar model to study the electron-PCP correlation related effects and to evaluate the utility of the recently designed electron-PCP correlation functionals.

To construct the proper model, one of the two electrons in the harmonium model is replaced with a particle having a unit of the positive charge and an arbitrary mass (equal to or larger than the electron’s mass). Our previous studies indeed revealed that in real molecules, on average, a single electron surrounds a proton/muon thus the model hopefully must simulate hydrogen (H)/muonium (Mu) atom bonded within a molecule \cite{goli_DecipheringChemicalNature_2014,goli_TopologicalAIMAnalyses_2015,goli_WherePlacePositive_2015,goli_DevelopingEffectiveElectroniconly_2018}. The corresponding Hamiltonian of the model, hereafter called the H/Mu atom in the double harmonic traps, and abbreviated as H/Mu-DHT, is as follows:
\begin{align}\label{eq:1}
\hat{H}=& {{\hat{T}}_\mathrm{e}}+{{\hat{T}}_\mathrm{PCP}}+{{\hat{V}}_\mathrm{e\textrm{-}PCP}}+\nu_\mathrm{e}^\mathrm{ext}\left( {{{\vec{r}}}_\mathrm{e}} \right)+\nu_\mathrm{PCP}^\mathrm{ext}\left( {{{\vec{r}}}_\mathrm{PCP}} \right) \nonumber \\
=& -\frac{1}{2} \nabla _\mathrm{e}^{2} -\frac{1}{2{{m}_\mathrm{PCP}}} \nabla _\mathrm{PCP}^{2} - \frac{1}{\left| {{{\vec{r}}}_\mathrm{e}}-{{{\vec{r}}}_\mathrm{PCP}} \right|} \nonumber \\  
&+ \frac{1}{2} {{k}_\mathrm{e}}r_\mathrm{e}^{2} + \frac{1}{2} {{k}_\mathrm{PCP}}r_\mathrm{PCP}^{2}.
\end{align}
Note that both in this equation and throughout the rest of the text, all equations and numerical data are given in the atomic units unless stated otherwise. The parameters of the model are the mass of the PCP, $m_\mathrm{PCP}$, and the force constants of the two traps, one for the electron, ${k}_\mathrm{e}$, and the other for the PCP, ${k}_\mathrm{PCP}$, sharing a single center in space. {Each trap acts as the external potential for the electron, $\nu_\mathrm{e}^\mathrm{ext}$, or the PCP, $\nu_\mathrm{PCP}^\mathrm{ext}$.} A special case of this model, ${k}_\mathrm{e}=0$, has been considered recently \cite{stetzler_ComparisonBornOppenheimer_2023}, but since we are interested in a separable Hamiltonian (\textit{vide infra}), we further assume that the frequency of oscillations is equal for the two traps: ${k}_{i}=m_{i}\omega^2$. Let us stress at this stage of development that the model can be used to simulate various physical systems that some are beyond the intent of the present study including the trapped atoms \cite{fernandez_ConfinedHydrogenAtom_2010,fernandez_VariationalTreatmentConfined_2010,fernandez_VariationalApproachConfined_2012,randazzo_EndohedrallyConfinedHydrogen_2016}, and the electron-hole pair \cite{kayanuma_WannierExcitonMicrocrystals_1986,nair_QuantumSizeEffects_1987,kayanuma_QuantumsizeEffectsInteracting_1988,kayanuma_IncompleteConfinementElectrons_1990a,elward_InvestigationElectronHole_2012,elward_CalculationElectronholeRecombination_2012}. The Hamiltonian is transformed by employing the center of mass, $\vec{R}={( {{{\vec{r}}}_\mathrm{e}}+{{m}_\mathrm{PCP}}{{{\vec{r}}}_\mathrm{PCP}} )}/{M}$, and the relative, $\vec{r}={{{{\vec{r}}}_\mathrm{e}}-{{{\vec{r}}}_\mathrm{PCP}}}$, variables into two independent Hamiltonians without any coupling term:            
\begin{align}\label{eq:2}
\hat{H}&= \hat{H}_{R}+\hat{H}_{r}, \nonumber \\ 
{{\hat{H}}_{R}}&= -\frac{1}{2M} \nabla _{R}^{2} + \frac{1}{2} M {\omega }^{2} {{R}^{2}}, \nonumber \\ 
{{\hat{H}}_{r}}&= -\frac{1}{2\mu } \nabla _{r}^{2}+ \frac{1}{2} \mu {\omega }^{2} {{r}^{2}}- \frac{1}{r}.
\end{align}
The center of mass Hamiltonian, $\hat{H}_{R}$, describes a pseudo-particle with the total mass $M=1+m_\mathrm{PCP}$, in a harmonic trap which is a textbook example of an analytically solvable model \cite{mcintyre_QuantumMechanicsParadigms_2012}. The relative motion Hamiltonian, $\hat{H}_{r}$, describes a pseudo-particle with the reduced mass $\mu=m_\mathrm{PCP}/(1+m_\mathrm{PCP})$, experiencing a harmonic potential and a Coulombic attraction. The mathematical procedures used to derive the analytical solutions of the relative motion Hamiltonian of the harmonium atom do not provide us with the analytical solution for the ground state of $\hat{H}_{r}$ \cite{taut_TwoElectronsExternal_1993,ushveridze_QuasiExactlySolvableModels_1994,karwowski_Harmonium_2004,karwowski_BiconfluentHeunEquation_2014}. Accordingly, for different sets of the model’s parameters, we derived the ground state wavefunctions of the relative motion from high-precision numerical solutions through the finite element method.

To test the physical relevance of the model, the mass and the frequency of oscillation of the model must be fixed to values that make comparison to real physical systems feasible. Since in this study, we are interested in the correlations between localized PCPs and electrons, the mass of the PCP was fixed at the masses of proton, $m_\mathrm{proton} \approx 1836$, and muon, $m_\mathrm{muon}\approx 207$. Fixing the frequency of oscillation and concomitant force constants is a more delicate issue, since, based on the separability condition of the Hamiltonian, the frequency of oscillation must be equal for the electron and the proton/muon. In fact, the force constants may be viewed as a simple representation of how the other nuclei and electrons collectively affect the target electron and the proton/muon in a real molecule. It seems reasonable to assume that the target electron is mainly influenced by the nearby proton/muon and is least affected by other nuclei/electrons and this justifies the assumption ${k}_\mathrm{e}=0$ \cite{stetzler_ComparisonBornOppenheimer_2023}. However, in principle, any other value for the force constant that makes it small relative to the Coulomb term in the vicinity of the center of the trap is also physically a reasonable choice. We fixed the frequency of oscillation by taking into account the typical known values of the zero-point vibrational energies of proton/muon in real molecules, ${\omega}_\mathrm{proton}=0.01$ and ${\omega}_\mathrm{muon}=0.02$ \cite{lin-vien_HandbookInfraredRaman_1991,jayasooriya_StrategyMeasurementVibrations_2004}. With this strategy, since the force constants are scaled with the particle’s mass, ${k}_{i}=m_{i}\omega^2$, the relative smallness of ${k}_\mathrm{e}$ is guaranteed. Nevertheless, the main results gained in this study are not sensitive to the numerical choice of $\omega$ and remain valid in a broad range of oscillation frequencies.  

\subsection{The overlocalization of the uncorrelated one-proton/muon densities}

\begin{figure}[t]
\includegraphics[width=\columnwidth]{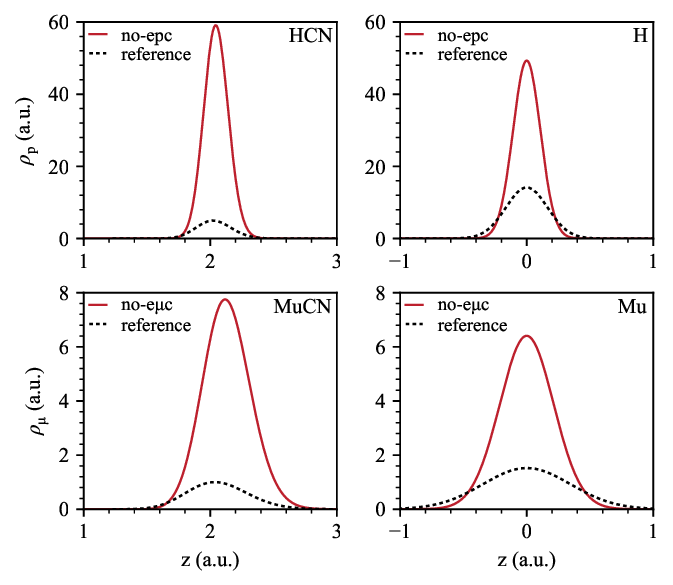}%
\caption{The left-hand panels depict $\rho_\mathrm{p/\upmu }$ for HCN and MuCN, while the right-hand panels depict the same densities for H/Mu-DHT. The reference $\rho_\mathrm{p/\upmu }$ of HCN and MuCN, depicted as dashed curves, were derived in a previous study from the double-adiabatic approximation \cite{goli_TwocomponentDensityFunctional_2022}. The reference $\rho_\mathrm{p/\upmu }$ for the model, depicted as dashed curves, is obtained from the computed exact wavefunction. The uncorrelated one-particle densities of the model, the solid red curves, are obtained at the MCHF/[7s7p7d/7s7p7d] level while for HCN and MuCN they are derived at the B3LYP/pc-2//no-epc/14s14p14d and B3LYP/pc-2//no-\emc/14s14p14d levels (for more details see Appendix \ref{sec:a2}). In the case of the model, all the one-particle densities are isotropic and the center of the coordinate system is placed at the joint center of the traps. For the real molecules, the one-particle densities are anisotropic and the axis used to depict the densities goes through the maximum of $\rho_\mathrm{p/\upmu }$ and the clamped carbon while the latter is located at the center of the coordinate system.}
\label{fig:1}
\end{figure}

\begin{figure}[t]
\includegraphics[width=\columnwidth]{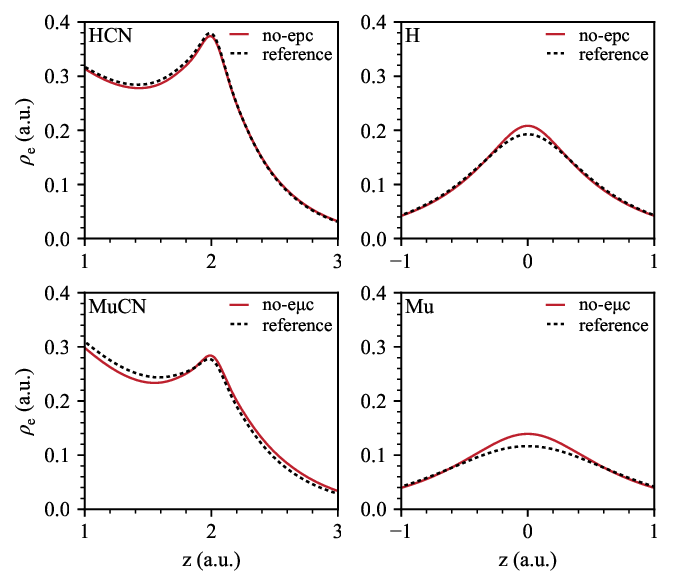}%
\caption{The left-hand panels depict $\rho_\mathrm{e}$ for HCN and MuCN, while the right-hand panels depict the same densities for H/Mu-DHT. The reference $\rho_\mathrm{e}$ of HCN and MuCN, depicted as dashed curves, are computed at the B3LYP/pc-2//epc17-1/14s14p14d and B3LYP/pc-2//\emc-1/14s14p14d levels, respectively, while the uncorrelated densities are derived at the B3LYP/pc-2//no-epc/14s14p14d and B3LYP/pc-2//no-\emc/14s14p14d levels, respectively (for more details see Appendix \ref{sec:a2}). The reference densities for the model are obtained from the computed exact wavefunction, whereas the uncorrelated densities are derived at the MCHF/[7s7p7d/7s7p7d] level. In the case of the model, all $\rho_\mathrm{e}$ are isotropic and the center of the coordinate system is placed at the joint center of the traps. For the real molecules, $\rho_\mathrm{e}$ are anisotropic and the axis used to depict the densities goes through the maximum of $\rho_\mathrm{p/\upmu }$ and the clamped carbon nucleus where the latter is located at the center of the coordinate system.}
\label{fig:2}
\end{figure}

In the first step, the exact and the mean-field derived one-proton/muon densities were studied to see whether the previously mentioned overlocalization is taking place also in the model. This is a crucial characteristic without which the model would not be proper for the quality evaluation of the electron-PCP correlation functionals. The one-particle densities in the original variables are easily deducible from the exact ground state wavefunction of the model: ${{\rho }_\mathrm{e}}({{{\vec{r}}}_\mathrm{e}})=\int{d{{{\vec{r}}}_\mathrm{p/\upmu }}}{{| {{\Psi }_\mathrm{exact}}( {{{\vec{r}}}_\mathrm{e}},{{{\vec{r}}}_\mathrm{p/\upmu }}) |}^{2}}$, ${{\rho }_\mathrm{p/\upmu }}({{\vec{r}}}_\mathrm{p/\upmu } )=\int{d{{{\vec{r}}}_\mathrm{e}}}{{| {{\Psi }_\mathrm{exact}}( {{{\vec{r}}}_\mathrm{e}},{{{\vec{r}}}_\mathrm{p/\upmu }} ) |}^{2}}$. Note that since Eq. (\ref{eq:2}) is solved in practice, the ground state wavefunction is derived according to the secondary variables, $\Psi(\vec{r},\vec{R})$. Therefore, before doing the integration, a “back transformation” must be done to the original variables where its mathematical details have been disclosed previously by Laufer and Krieger and are not reiterated herein \cite{laufer_TestDensityfunctionalApproximations_1986}. The derived one-particle densities are the “reference” densities of the model for all subsequent comparisons.    

To check the presence of the overlocalization in {${\rho }_\mathrm{p/\upmu }$} of the model, as discussed previously, we need to have also an uncorrelated description of the electron and the proton/muon. Accordingly, we tried to find the best variational Hartree-product solution for the ground state of the model, $\Psi _\mathrm{uncorrelated} ( {{{\vec{r}}}_\mathrm{e}},{{{\vec{r}}}_\mathrm{p/\upmu }} )={{\phi }_\mathrm{e}} ( {{{\vec{r}}}_\mathrm{e}} ){{\phi }_\mathrm{p/\upmu }}( {{{\vec{r}}}_\mathrm{p/\upmu }} )$, which is formally equivalent to the MCHF wavefunction for a system composed of an electron and a proton/muon. This was done by expanding ${\phi }_\mathrm{e}$ and ${\phi }_\mathrm{p/\upmu}$ in a series of seven s-, p- and d-type Gaussian functions denoted as [7s7p7d/7s7p7d] basis set; details of the basis set are given in Appendix \ref{sec:a1}. It was found during the numerical tests that the MCHF/[7s7p7d/7s7p7d] level yields numerical results near the infinite basis limit for the practical applications intended in this study. From the derived uncorrelated wavefunction, the uncorrelated one-particle densities are easily deducible: ${\rho }_\mathrm{e}=|{\phi }_\mathrm{e}|^2$ and ${\rho }_\mathrm{p/\upmu }=|\phi _\mathrm{p/\upmu}|^2$. For comparison purposes, the same one-particle densities were also computed for hydrogen/muonium cyanide, XCN (X=H, Mu), where the latter is the simplest muonic molecule considered in our previous MCDFT study \cite{goli_TwocomponentDensityFunctional_2022}. The details of the used computational levels to deduce the “reference”/exact and uncorrelated one-particle densities may be found in our previous paper \cite{goli_TwocomponentDensityFunctional_2022}, and a brief survey is also given in Appendix \ref{sec:a2}. 

Figures \ref{fig:1} and \ref{fig:2} depict 1D slices of the one-particle densities computed from the exact and uncorrelated wavefunctions for the model and the XCN molecules. In the case of the model, since the one-particle densities are isotropic, the axis used to depict the figures is arbitrary and goes through the joint center of the traps, whereas for XCN species, they are anisotropic and the axis {passes} through the maximum of ${\rho }_\mathrm{p/\upmu }$ and the clamped carbon and nitrogen nuclei is used for depiction \cite{goli_TwocomponentDensityFunctional_2022}. A quick glance at Fig. \ref{fig:1} reveals unequivocally the expected overlocalization of the uncorrelated ${\rho }_\mathrm{p/\upmu }$ in comparison to the reference densities in both real and model systems, whereas no such extreme overlocalization is observable in the case of ${\rho }_\mathrm{e}$ in Fig. \ref{fig:2}.

We conclude that the model, in contrast to its seemingly profound simplicity, reveals the same “pathological behavior” of the uncorrelated ${\rho }_\mathrm{p/\upmu}$ observable in the case of the real systems. In this regard, it can be used as a “laboratory” to study this pathological behavior in detail and to verify how a properly designed correlated wavefunction, or an electron-proton/muon correlation functional, may remedy the overlocalization.  

\subsection{The two-component KS inversion}\label{sec:2c}

As discussed in the introduction, the fundamental theorems of the MCDFT as well as the MCKS equations have been derived for the general MC systems long ago (for a compressed review see Sec. 9.6 in Ref. \cite{parr_DensityFunctionalTheoryAtoms_1994}). By the way, in this subsection, we briefly review the specific KS system of the model and also the inversion process upon which the exact electron-proton/muon correlational potential is derived. Numerical comparisons with approximate correlation functionals are considered in the next section.

The MC Hohenberg-Kohn theorems and the generalized Levy’s constraint search imply the following for finding the ground state energy of the model based on Eq. (\ref{eq:1}):
\begin{widetext}
\begin{align}\label{eq:3}
{{E}_\mathrm{ground}}=&\underset{{{\rho }_\mathrm{e}},{{\rho }_\mathrm{p/\upmu}}}{\mathop{\mathrm{min}}}\,E\left[ {{\rho }_\mathrm{e}},{{\rho }_\mathrm{p/\upmu }} \right], \nonumber \\ 
E\left[ {{\rho }_\mathrm{e}},{{\rho }_\mathrm{p/\upmu }} \right]=&F\left[ {{\rho }_\mathrm{e}},{{\rho }_\mathrm{p/\upmu }} \right]+\int{d{{{\vec{r}}}_\mathrm{e}} \; \nu_\mathrm{e}^\mathrm{ext}\left( {{{\vec{r}}}_\mathrm{e}} \right){{\rho }_\mathrm{e}}\left( {{{\vec{r}}}_\mathrm{e}} \right)} +\int{d{{{\vec{r}}}_\mathrm{p/\upmu }} \; \nu_\mathrm{p/\upmu }^\mathrm{ext}\left( {{{\vec{r}}}_\mathrm{p/\upmu }} \right){{\rho }_\mathrm{p/\upmu }}\left( {{{\vec{r}}}_\mathrm{p/\upmu }} \right)}, \nonumber \\ 
F\left[ {{\rho }_\mathrm{e}},{{\rho }_\mathrm{p/\upmu }} \right]  =&\underset{\Psi_\mathrm{e,p/\upmu} \to {{\rho }_\mathrm{e}},{{\rho }_\mathrm{p/\upmu }}}{\mathop{\mathrm{min}}} \left\langle {{\Psi }_\mathrm{e,p/\upmu}}\left| {{{\hat{T}}}_\mathrm{e}}+{{{\hat{T}}}_\mathrm{p/\upmu }}+{{{\hat{V}}}_\mathrm{e\textrm{-}(p/\upmu)}} \right|{{\Psi }_\mathrm{e,p/\upmu}} \right\rangle \nonumber \\ 
=&{{T}_\mathrm{e}}\left[ {{\rho }_\mathrm{e}},{{\rho }_\mathrm{p/\upmu }} \right]+{{T}_\mathrm{p/\upmu }}\left[ {{\rho }_\mathrm{e}},{{\rho }_\mathrm{p/\upmu }} \right] +{{V}_\mathrm{e\textrm{-}(p/\upmu)}}\left[ {{\rho }_\mathrm{e}},{{\rho }_\mathrm{p/\upmu }} \right].
\end{align}
% \end{widetext}
$F [ {{\rho }_\mathrm{e}},{{\rho }_\mathrm{p/\upmu }} ]$ is the Hohenberg-Kohn universal functional of the model which is a functional of ${\rho }_\mathrm{e}$ and ${\rho }_\mathrm{p/\upmu }$, independent from $\nu_\mathrm{e}^\mathrm{ext}$ and $\nu_\mathrm{p/\upmu }^\mathrm{ext}$. Assuming there is a non-interacting two-particle reference KS system for the model, reproducing the exact one-particle densities, the universal functional may be rewritten as follows:
% \begin{widetext}
\begin{align}\label{eq:4}
F\left[ {{\rho }_\mathrm{e}},{{\rho }_\mathrm{p/\upmu }} \right]=&T_\mathrm{e}^{s}\left[ {{\rho }_\mathrm{e}} \right]+T_{p/\mu }^{s}\left[ {{\rho }_\mathrm{p/\upmu }} \right] +{{J}_\mathrm{e\textrm{-}(p/\upmu) }}\left[ {{\rho }_\mathrm{e}},{{\rho }_\mathrm{p/\upmu }} \right] +{{E}_\mathrm{e\left( p/\upmu  \right)c}}\left[ {{\rho }_\mathrm{e}},{{\rho }_\mathrm{p/\upmu }} \right], \nonumber \\ 
{{E}_\mathrm{e\left( p/\upmu  \right)c}}\left[ {{\rho }_\mathrm{e}},{{\rho }_\mathrm{p/\upmu }} \right]=&\left( {{T}_\mathrm{e}}\left[ {{\rho }_\mathrm{e}},{{\rho }_\mathrm{p/\upmu }} \right]-T_\mathrm{e}^{s}\left[ {{\rho }_\mathrm{e}} \right] \right) +\left( {{T}_\mathrm{p/\upmu }}\left[ {{\rho }_\mathrm{e}},{{\rho }_\mathrm{p/\upmu }} \right]-T_\mathrm{p/\upmu }^{s}\left[ {{\rho }_\mathrm{p/\upmu }} \right] \right) \nonumber \\
&+\left( {{V}_\mathrm{e\textrm{-}(p/\upmu) }}\left[ {{\rho }_\mathrm{e}},{{\rho }_\mathrm{p/\upmu }} \right]-{{J}_\mathrm{e\textrm{-}(p/\upmu) }}\left[ {{\rho }_\mathrm{e}},{{\rho }_\mathrm{p/\upmu }} \right] \right).
\end{align}
% \end{widetext}
In these equations $T_\mathrm{e}^{s}[ {{\rho }_\mathrm{e}} ]=({-1}/{2})\langle \phi _\mathrm{e}^\mathrm{KS} | \nabla _\mathrm{e}^{2} |\phi _\mathrm{e}^\mathrm{KS} \rangle $ and $T_\mathrm{p/\upmu }^{s} [ {{\rho }_\mathrm{p/\upmu }} ]= ( {-1}/{2{{m}_\mathrm{p/\upmu }}}  ) \langle \phi _\mathrm{p/\upmu }^\mathrm{KS} | \nabla _\mathrm{p/\upmu }^{2}  |\phi _\mathrm{p/\upmu }^\mathrm{KS}  \rangle $ are the KS non-interacting kinetic energies of the electron and the proton/muon while ${{J}_\mathrm{e\textrm{-}(p/\upmu) }} [ {{\rho }_\mathrm{e}},{{\rho }_\mathrm{p/\upmu }} ]=-\int{d{{{\vec{r}}}_\mathrm{e}}\int{d{{{\vec{r}}}_\mathrm{p/\upmu }}}}\frac{{{\rho }_\mathrm{e}} ( {{{\vec{r}}}_\mathrm{e}} ){{\rho }_\mathrm{p/\upmu }} ( {{{\vec{r}}}_\mathrm{p/\upmu }} )}{| {{{\vec{r}}}_\mathrm{e}}-{{{\vec{r}}}_\mathrm{p/\upmu }} |}$ is the classical Coulomb interaction energy, sometimes also called the Hartree term. Neglecting the spin, the KS wavefunction of the model is the product of the KS spatial orbitals: $\Psi _{KS}( {{{\vec{r}}}_\mathrm{e}},{{{\vec{r}}}_\mathrm{p/\upmu }} )=\phi _\mathrm{e}^\mathrm{KS}( {{{\vec{r}}}_\mathrm{e}} )\phi _\mathrm{p/\upmu }^\mathrm{KS} ( {{{\vec{r}}}_\mathrm{p/\upmu }} )$, where: ${\rho }_\mathrm{e}=|{\phi }_\mathrm{e}^\mathrm{KS}|^2$ and ${\rho }_\mathrm{p/\upmu }=|\phi _\mathrm{p/\upmu}^\mathrm{KS}|^2$ are the one-particle densities of the model. The only unknown is the functional form of the electron-proton/muon correlation functional, ${{E}_\mathrm{e( p/\upmu  )c}}[ {{\rho }_\mathrm{e}},{{\rho }_\mathrm{p/\upmu }} ]$. Upon the variation of the energy functional, Eq. (\ref{eq:3}), with respect to the KS spatial orbitals, the following set of coupled KS equations are derived:
% \begin{widetext}
\begin{align}\label{eq:5}
&\left\{  \begin{array}{l}
   \left( -\frac{1}{2} \nabla _\mathrm{e}^{2}+\nu_\mathrm{e}^\mathrm{KS}\left( {{{\vec{r}}}_\mathrm{e}} \right) \right)\phi _\mathrm{e}^\mathrm{KS}\left( {{{\vec{r}}}_\mathrm{e}} \right)=\varepsilon _\mathrm{e}^\mathrm{KS}\phi _\mathrm{e}^\mathrm{KS}\left( {{{\vec{r}}}_\mathrm{e}} \right)  \\
   \left( -\frac{1}{2{{m}_\mathrm{p/\upmu }}} \nabla _\mathrm{p/\upmu }^{2}+\nu_\mathrm{p/\upmu }^\mathrm{KS}\left( {{{\vec{r}}}_\mathrm{p/\upmu }} \right) \right)\phi _\mathrm{p/\upmu }^\mathrm{KS}\left( {{{\vec{r}}}_\mathrm{p/\upmu }} \right)=\varepsilon _\mathrm{p/\upmu }^\mathrm{KS}\phi _\mathrm{p/\upmu }^\mathrm{KS}\left( {{{\vec{r}}}_\mathrm{p/\upmu }} \right), \\
\end{array} \right. \nonumber \\ 
&\left\{  \begin{array}{l}
   \nu_\mathrm{e}^\mathrm{KS}\left( {{{\vec{r}}}_\mathrm{e}} \right)=\nu_\mathrm{e}^\mathrm{ext}\left( {{{\vec{r}}}_\mathrm{e}} \right)+\nu_\mathrm{e}^{J}\left( {{{\vec{r}}}_\mathrm{e}} \right)+\nu_\mathrm{e}^\mathrm{e\left( p/\upmu  \right)c}\left( {{{\vec{r}}}_\mathrm{e}} \right)  \\
   \nu_\mathrm{p/\upmu }^\mathrm{KS}\left( {{{\vec{r}}}_\mathrm{p/\upmu }} \right)=\nu_\mathrm{p/\upmu }^\mathrm{ext}\left( {{{\vec{r}}}_\mathrm{p/\upmu }} \right)+\nu_\mathrm{p/\upmu }^{J}\left( {{{\vec{r}}}_\mathrm{p/\upmu }} \right)+\nu_\mathrm{p/\upmu }^\mathrm{e\left( p/\upmu  \right)c}\left( {{{\vec{r}}}_\mathrm{p/\upmu }} \right).  \\
\end{array} \right.
\end{align}
% \end{widetext}
In these equations $\nu_\mathrm{e}^\mathrm{KS}( {{{\vec{r}}}_\mathrm{e}} )$ and $\nu_\mathrm{p/\upmu }^\mathrm{KS}( {{{\vec{r}}}_\mathrm{p/\upmu }} )$ are the effective KS electronic and protonic/muonic potentials, respectively. Their components, apart from the external potentials, are $\nu_\mathrm{e}^{J} ( {{{\vec{r}}}_\mathrm{e}} )=\frac{\delta {{J}_\mathrm{e\textrm{-}(p/\upmu) }}}{\delta {{\rho }_\mathrm{e}}}=-\int{d{{{\vec{r}}}_\mathrm{p/\upmu }}}\frac{{{\rho }_\mathrm{p/\upmu }}( {{{\vec{r}}}_\mathrm{p/\upmu }} )}{| {{{\vec{r}}}_\mathrm{e}}-{{{\vec{r}}}_\mathrm{p/\upmu }} |}$ and $\nu_\mathrm{p/\upmu }^{J}( {{{\vec{r}}}_\mathrm{p/\upmu }} )=\frac{\delta {{J}_\mathrm{e\textrm{-}(p/\upmu) }}}{\delta {{\rho }_\mathrm{p/\upmu }}}=-\int{d{{{\vec{r}}}_\mathrm{e}}}\frac{{{\rho }_\mathrm{e}}( {{{\vec{r}}}_\mathrm{e}} )}{| {{{\vec{r}}}_\mathrm{e}}-{{{\vec{r}}}_\mathrm{p/\upmu }} |}$, as the potentials emerging from the Hartree term, and, $\nu_\mathrm{e}^\mathrm{e ( p/\upmu  )c} ( {{{\vec{r}}}_\mathrm{e}} )=\frac{\delta {{E}_\mathrm{e( p/\upmu )c}}}{\delta {{\rho }_\mathrm{e}}}$ and $\nu_{p/\upmu }^\mathrm{e( p/\upmu)c}( {{{\vec{r}}}_\mathrm{p/\upmu }} )=\frac{\delta {{E}_\mathrm{e( p/\upmu  )c}}}{\delta {{\rho }_\mathrm{p/\upmu }}}$, as the electronic and protonic/muonic correlation potentials, respectively. 

Since the exact functional form of ${{E}_\mathrm{e( p/\upmu  )c}}[ {{\rho }_\mathrm{e}},{{\rho }_\mathrm{p/\upmu }} ]$ is unknown, in order to start solving the KS equations, an approximate functional form must be used to deduce the corresponding correlation potentials. Such approximate functionals are considered in the next section, but since ${\Psi }_\mathrm{exact}( {{{\vec{r}}}_\mathrm{e}},{{{\vec{r}}}_\mathrm{p/\upmu }})$ and the concomitant exact ${{\rho }_\mathrm{e}}$ and ${{\rho }_\mathrm{p/\upmu }}$ are known, depicted in Fig. \ref{fig:3}, the exact correlations potentials are also deducible from an inversion process similar to that employed previously in the case of the harmonium model \cite{laufer_TestDensityfunctionalApproximations_1986,kais_DensityFunctionalsDimensional_1993,filippi_ComparisonExactApproximate_1994}. The relevant expressions, derived from the coupled KS equations, are the following: 
% \begin{widetext}
\begin{align}\label{eq:6}
    \left\{ \begin{array}{l}
   \nu_\mathrm{e}^\mathrm{e\left( p/\upmu  \right)c}\left( {{{\vec{r}}}_\mathrm{e}} \right)=\varepsilon _\mathrm{e}^\mathrm{KS}-\nu_\mathrm{e}^\mathrm{ext}\left( {{{\vec{r}}}_\mathrm{e}} \right)-\nu_\mathrm{e}^{J}\left( {{{\vec{r}}}_\mathrm{e}} \right)+ \frac{\nabla _\mathrm{e}^{2} \phi _\mathrm{e}^\mathrm{KS}\left( {{{\vec{r}}}_\mathrm{e}} \right) }{2\phi _\mathrm{e}^\mathrm{KS}\left( {{{\vec{r}}}_\mathrm{e}} \right)}  \\
   \nu_\mathrm{p/\upmu }^\mathrm{e\left( p/\upmu  \right)c}\left( {{{\vec{r}}}_\mathrm{p/\upmu }} \right)=\varepsilon _\mathrm{p/\upmu }^\mathrm{KS}-\nu_\mathrm{p/\upmu }^\mathrm{ext}\left( {{{\vec{r}}}_\mathrm{p/\upmu }} \right)-\nu_\mathrm{p/\upmu }^{J}\left( {{{\vec{r}}}_\mathrm{p/\upmu }} \right)+ \frac{\nabla _\mathrm{p/\upmu }^{2} \phi _\mathrm{p/\upmu }^\mathrm{KS}\left( {{{\vec{r}}}_\mathrm{p/\upmu }} \right) }{2{{m}_\mathrm{p/\upmu }} \phi _\mathrm{p/\upmu }^\mathrm{KS}\left( {{{\vec{r}}}_\mathrm{p/\upmu }} \right)}.  \\
    \end{array} \right.
\end{align}
\end{widetext}
Comparison between the exact and an approximate correlation potential is probably the most stringent quality measure of an approximate functional as also demonstrated in the case of the electronic exchange-correlation functionals \cite{laufer_TestDensityfunctionalApproximations_1986,kais_DensityFunctionalsDimensional_1993,filippi_ComparisonExactApproximate_1994,umrigar_AccurateExchangecorrelationPotentials_1994,gritsenko_MolecularKohnShamExchangecorrelation_1995,gritsenko_MolecularExchangeCorrelation_1996,kananenka_EfficientConstructionExchange_2013,ryabinkin_ReductionElectronicWave_2015,kanungo_ExactExchangecorrelationPotentials_2019,kumar_AccurateEffectivePotential_2020,kanungo_ComparisonExactModel_2021}. Let us now turn to the numerical application of the formalism to the model.      

Since $| \phi _\mathrm{e}^\mathrm{KS} |=\sqrt{{{\rho }_\mathrm{e}}}$ and $| \phi _\mathrm{p/\upmu }^\mathrm{KS} |=\sqrt{{{\rho }_\mathrm{p/\upmu }}}$, apart from the phase, $\phi _\mathrm{e}^\mathrm{KS}$ and $\phi _\mathrm{p/\upmu }^\mathrm{KS}$ are available, and the numerical values of all energy components except for ${{E}_\mathrm{e( p/\upmu  )c}}$, namely, $T_\mathrm{e}^{s}$, $T_\mathrm{p/\upmu}^{s}$, ${J}_\mathrm{e\textrm{-}(p/\upmu) }$, $\langle \nu_\mathrm{e }^\mathrm{ext} \rangle$, $\langle \nu_\mathrm{p/\upmu }^\mathrm{ext} \rangle$, are easily computable. The eigenvalue {problems} of the Hamiltonians given in Eq. (\ref{eq:2}) are solved exactly and $E_\mathrm{exact}=E_R+E_r$ is numerically known, thus, ${{E}_\mathrm{e( p/\upmu  )c}}$ is computed as follows: ${{E}_\mathrm{e( p/\upmu  )c}}=E_\mathrm{exact}-(T_\mathrm{e}^{s}+ T_\mathrm{p/\upmu}^{s}+{J}_\mathrm{e\textrm{-}(p/\upmu) }+\langle \nu_\mathrm{e }^\mathrm{ext} \rangle+\langle \nu_\mathrm{p/\upmu }^\mathrm{ext} \rangle)$. Alternatively, since the expectation values of the operators composing the original Hamiltonian, Eq. (\ref{eq:1}), are all known, it is feasible to compute ${{E}_\mathrm{e( p/\upmu  )c}}$ directly using Eq. (\ref{eq:4}); both methods yield the same numerical value. 
	
\begin{figure}[t]
\includegraphics[width=\columnwidth]{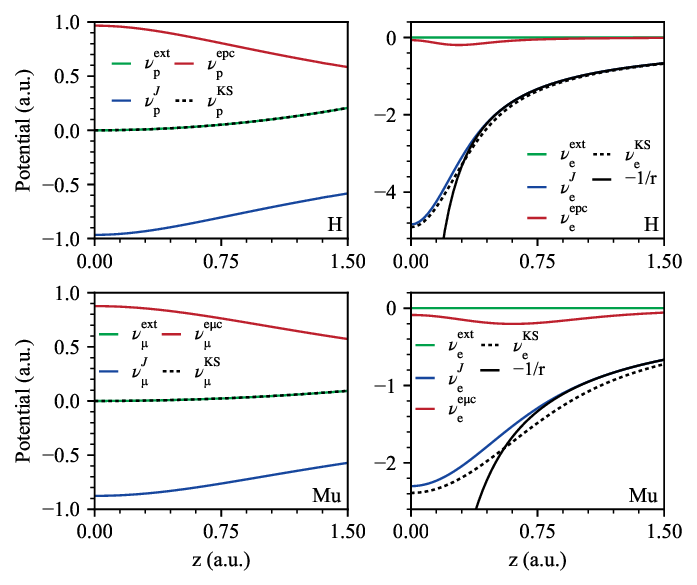}%
\caption{The reference $\nu_\mathrm{e}^\mathrm{KS}$ and $\nu_\mathrm{p/\upmu}^\mathrm{KS}$ and their components $\nu_\mathrm{e}^\mathrm{ext}$, $\nu_\mathrm{e}^{J}$, $\nu_\mathrm{e}^\mathrm{e(p/\upmu)c}$, and $\nu_\mathrm{p/\upmu}^\mathrm{ext}$, $\nu_\mathrm{p/\upmu}^J$, $\nu_\mathrm{p/\upmu}^\mathrm{e(p/\upmu)c}$, respectively, for H/Mu-DHT (see {Eq. (\ref{eq:5})} for more details). The Coulomb's law, the solid black line, is also given in the right-hand panels for comparison. Since all the potentials are isotropic, the direction of the z-axis is arbitrary and the center of the coordinate system is placed at the joint center of the double harmonic traps.}
\label{fig:4}
\end{figure}

All total energies and their components are gathered for H/Mu-DHT in Table \ref{tab:1}. Most notable is the twice larger ${{E}_\mathrm{e\upmu c}}$ in comparison to ${{E}_\mathrm{epc}}$, consistent with the known fact that the correlation energy is inversely related to the mass of the PCP. In order to derive the KS orbital energies, $\varepsilon _\mathrm{e}^\mathrm{KS}$ and $\varepsilon _\mathrm{p/\upmu }^\mathrm{KS}$, the known trick of imposing the following asymptotic conditions was used: ${{\lim }_{| {{{\vec{r}}}_\mathrm{e}} |\to \infty }}\nu_\mathrm{e}^\mathrm{e(p/\upmu )c}( {{{\vec{r}}}_\mathrm{e}} )\to 0$ and ${{\lim }_{| {{{\vec{r}}}_\mathrm{p/\upmu }} |\to \infty }}\nu_\mathrm{p/\upmu }^\mathrm{e( p/\upmu )c}( {{{\vec{r}}}_\mathrm{p/\upmu }} )\to 0$ (for more details see Refs. \cite{laufer_TestDensityfunctionalApproximations_1986,kais_DensityFunctionalsDimensional_1993,filippi_ComparisonExactApproximate_1994}), and the numerical results are also given in Table \ref{tab:1}. Interestingly, the values of $\varepsilon _\mathrm{e}^\mathrm{KS}$ in both systems are very near the ground state energy of the free hydrogen atom, $\sim -0.5$, while those of $\varepsilon _\mathrm{p/\upmu }^\mathrm{KS}$ are practically equal to the zero-point energy of the harmonic traps, $\varepsilon _\mathrm{p}^\mathrm{KS} \sim \frac{3}{2}\omega_\mathrm{proton}=0.015$ and $\varepsilon _\mathrm{\upmu }^\mathrm{KS}\sim \frac{3}{2}\omega_\mathrm{muon}=0.03$. To have a “local” picture of the role of correlations, Fig. \ref{fig:4} depicts $\nu_\mathrm{e}^\mathrm{KS}$ and $\nu_\mathrm{p/\upmu }^\mathrm{KS}$ as well as their components individually as introduced in Eq. (\ref{eq:5}). It is evident that the role of the protonic/muonic correlation potentials is pivotal, eliminating the Hartree part of the potential, $\nu_\mathrm{p/\upmu }^\mathrm{e(p/\upmu)c}\approx -\nu_\mathrm{p/\upmu }^{J}$. Thus, to a very good approximation: $\nu_\mathrm{p/\upmu }^\mathrm{KS}\approx\nu_\mathrm{p/\upmu }^\mathrm{ext}$, which explains why the numerical values of $\varepsilon_\mathrm{p/\upmu }^\mathrm{KS}$ are equal to the zero-point energy of the traps. In contrast, the contribution of ${{\nu}_\mathrm{e}^\mathrm{e( p/\upmu  )c}}$ is marginal in shaping $\nu_\mathrm{e}^\mathrm{KS}$ and only modifies the dominant contribution of $\nu_\mathrm{e}^{J}$ slightly, thus, to a good approximation: $\nu_\mathrm{e}^\mathrm{KS}\approx \nu_\mathrm{e}^{J}$. As is evident from the figure, upon taking some distance from the joint center of the traps the Hartree potential quickly approaches the Coulomb law: $\nu_\mathrm{e}^{J}\sim -\frac{1}{r}$, and this may explain the origin of the numerical values of $\varepsilon_\mathrm{e}^\mathrm{KS}$. 
	
Based on these observations, we conclude that the model is capable of providing a clear understanding of the roles of the correlation potentials, however, as is discussed in the next section, the reproduction of these potentials through applying approximate functionals to the coupled KS equations is not an easy task.     

\begin{figure}[t]
\includegraphics[width=\columnwidth]{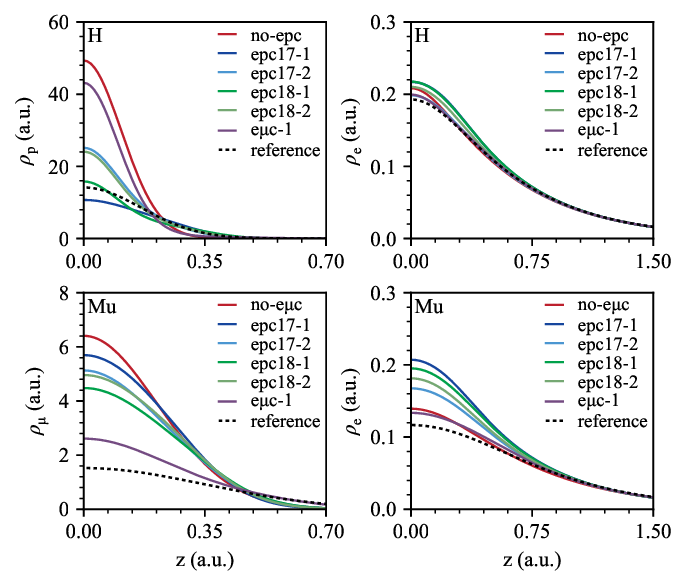}%
\caption{1D slices of the reference, functional- and  MCHF-derived $\rho_\mathrm{e}$ and $\rho_\mathrm{p/\upmu }$ for H/Mu-DHT. Since all the one-particle densities are isotropic, the direction of the z-axis is arbitrary and the center of the coordinate system is placed at the joint center of the double harmonic traps.}
\label{fig:3}
\end{figure}

\begin{table*}[t]
  \centering
  \caption{The exact/reference and approximate KS energy components computed for H/Mu-DHT (for the definition of each term see Sec. \ref{sec:2c}). The entry denoted as “reference” contains the exact results while the others are obtained with and without the considered electron-proton/muon correlation functionals.}
  \begin{ruledtabular}
    \begin{tabular}{c d{2.4}d{1.4}d{1.4}d{1.4}d{1.4}d{2.4}d{2.4}d{2.2}d{2.3}}
     Method     & \multicolumn{1}{c}{$E_\mathrm{ground}$} & \multicolumn{1}{c}{$T_\mathrm{e}^s$}    & \multicolumn{1}{c}{$\langle \nu_\mathrm{e}^\mathrm{ext} \rangle$}    & \multicolumn{1}{c}{$T_\mathrm{p/\upmu}^s$}    & \multicolumn{1}{c}{$\langle \nu_\mathrm{p/\upmu}^\mathrm{ext} \rangle$}    & \multicolumn{1}{c}{$J_\mathrm{e\textrm{-}(p/\upmu)}$}    & \multicolumn{1}{c}{$E_\mathrm{e(p/\upmu) c}$}   & \multicolumn{1}{c}{$\varepsilon_\mathrm{e}^\mathrm{KS}$}  & \multicolumn{1}{c}{$\varepsilon_\mathrm{p/\upmu}^\mathrm{KS}$}   \\ \hline
     & \multicolumn{9}{c}{H} \\ \cmidrule{2-10}
    no-epc/MCHF    & -0.4628 & 0.4561 & 0.0002 & 0.0171 & 0.0033 & -0.9393 &  0.0000 & -0.48 & -0.919 \\
    epc17-1   & -0.5112 & 0.4928 & 0.0001 & 0.0083 & 0.0075 & -0.9564 & -0.0636 & -0.50 & -0.978 \\
    epc17-2   & -0.4874 & 0.4606 & 0.0002 & 0.0094 & 0.0062 & -0.9323 & -0.0314 & -0.49 & -0.932 \\
    epc18-1   & -0.4971 & 0.4928 & 0.0001 & 0.0074 & 0.0085 & -0.9527 & -0.0532 & -0.50 & -0.959 \\
    epc18-2   & -0.4829 & 0.4751 & 0.0002 & 0.0092 & 0.0066 & -0.9443 & -0.0297 & -0.49 & -0.939 \\
    \emc-1    & -0.4786 & 0.4646 & 0.0002 & 0.0121 & 0.0077 & -0.9319 & -0.0314 & -0.49 & -0.922 \\
    reference & -0.4846 & 0.4649 & 0.0002 & 0.0075 & 0.0075 & -0.9315 & -0.0331 & -0.49 &  0.015 \\
          & \multicolumn{9}{c}{Mu} \\ \cmidrule{2-10}
    no-{\emc}/MCHF   & -0.4078 & 0.3892 & 0.0007 & 0.0383 & 0.0059 & -0.8419 &  0.0000 & -0.45 & -0.798 \\
    epc17-1   & -0.4690 & 0.4882 & 0.0006 & 0.0415 & 0.0055 & -0.9351 & -0.0696 & -0.50 & -0.942 \\
    epc17-2   & -0.4486 & 0.4378 & 0.0007 & 0.0360 & 0.0063 & -0.8855 & -0.0439 & -0.48 & -0.873 \\
    epc18-1   & -0.4628 & 0.4762 & 0.0006 & 0.0357 & 0.0064 & -0.9180 & -0.0637 & -0.50 & -0.916 \\
    epc18-2   & -0.4432 & 0.4515 & 0.0006 & 0.0381 & 0.0060 & -0.8995 & -0.0399 & -0.48 & -0.882 \\
    \emc-1    & -0.4547 & 0.4072 & 0.0007 & 0.0213 & 0.0108 & -0.8328 & -0.0619 & -0.47 & -0.821 \\
    reference & -0.4670 & 0.3944 & 0.0007 & 0.0151 & 0.0149 & -0.8035 & -0.0885 & -0.49 &  0.030 \\
    \end{tabular}%
    \end{ruledtabular}
  \label{tab:1}%
\end{table*}%

\section{Assessing the quality of the local electron-proton/muon correlation functionals}

\subsection{Applying approximate correlation functionals to the model}

In this section, some recently developed local electron-proton/muon correlation functionals are applied to the KS system of the model through the solution of the coupled KS equations. The resulting numerical data are compared to the reference solutions obtained in the previous section. Let us first briefly review the studied functionals and some computational details regarding the implementation of the coupled KS equations. 

There were several proposed electron-proton correlation functionals before the introduction of the epc series \cite{ito_FormulationNumericalApproach_2004,udagawa_IsotopeEffectPorphine_2006,imamura_ColleSalvettitypeCorrectionElectron_2008,imamura_ExtensionDensityFunctional_2009,udagawa_ElectronnucleusCorrelationFunctional_2014,udagawa_DevelopmentColleSalvettiType_2015}. However, to the best of our knowledge, none has systematically been evaluated through the standard self-consistent field (SCF) solution of the coupled KS equations on benchmark sets of molecules or crystals. Thus, in the present study we choose the local electron-proton correlation functionals from the epc series, namely, epc17-1 \cite{yang_DevelopmentPracticalMulticomponent_2017}, epc17-2 \cite{brorsen_MulticomponentDensityFunctional_2017}, epc18-1 and epc18-2 \cite{brorsen_AlternativeFormsTransferability_2018}, as well as \emc-1, as the electron-muon correlation functional \cite{goli_TwocomponentDensityFunctional_2022}; the functional forms are given in Appendix \ref{sec:a3}. All these functionals have been evaluated carefully in previous benchmark computational studies, and all are capable of remedying the overlocalization of the uncorrelated $\rho_\mathrm{p/\upmu}$. 

\begin{figure}[t]
\includegraphics[width=\columnwidth]{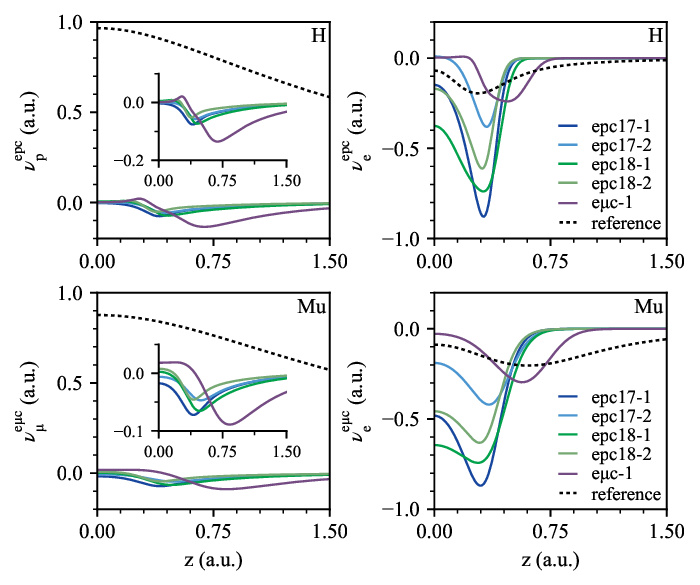}%
\caption{The reference and functional-derived {correlation potentials} for H/Mu-DHT. The reference correlation potentials are the exact results, depicted previously in Fig. \ref{fig:4}, while the others have been deduced using the considered electron-proton/muon correlation functionals. The potentials deduced for each functional are computed using $\rho_\mathrm{p/\upmu}$ {and $\rho_\mathrm{e}$} derived from the SCF solution of the coupled KS equations. {The insets are zoomed views of the functional-derived potentials.} Since all the potentials are isotropic, the direction of the z-axis is arbitrary and the center of the coordinate system is placed at the joint center of the double harmonic traps.}
\label{fig:5}
\end{figure}

\begin{figure}[t]
\includegraphics[width=\columnwidth]{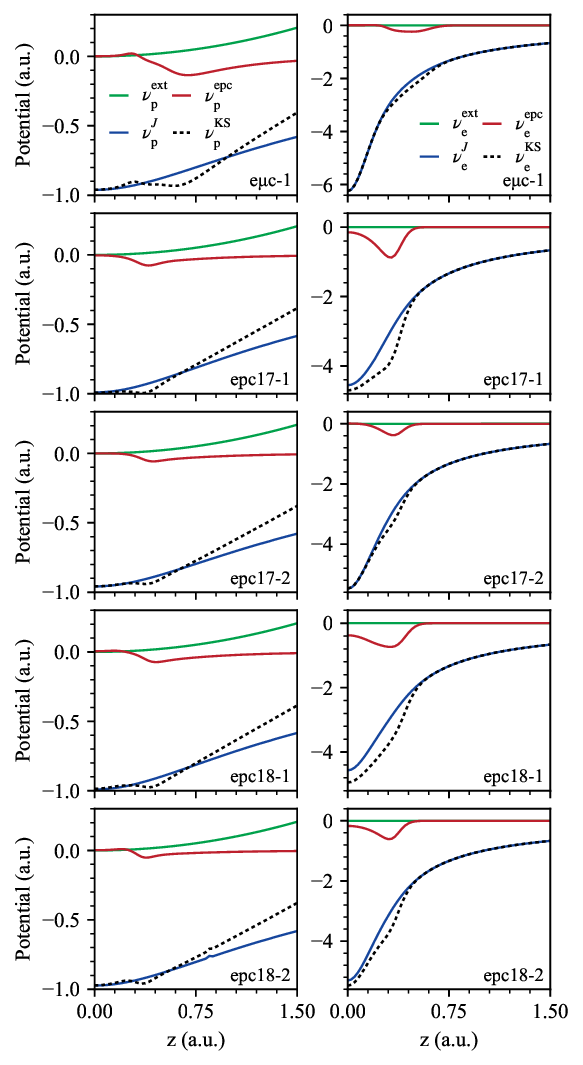}%
\caption{The functional-derived $\nu_\mathrm{e}^\mathrm{KS}$ and $\nu_\mathrm{p}^\mathrm{KS}$, and their components, $\nu_\mathrm{e}^\mathrm{ext}$, $\nu_\mathrm{e}^J$, $\nu_\mathrm{e}^\mathrm{epc}$, and $\nu_\mathrm{p}^\mathrm{ext}$, $\nu_\mathrm{p}^J$, $\nu_\mathrm{p}^\mathrm{epc}$, respectively, for H-DHT (see Eq. (\ref{eq:5}) for more details), deduced from the SCF-derived $\rho_\mathrm{e}$ and $\rho_\mathrm{p}$. {The protonic and electronic potentials are shown in the left- and right-hand panels, respectively.} Since all the potentials are isotropic, the direction of the z-axis is arbitrary and the center of the coordinate system is placed at the joint center of the double harmonic traps.}
\label{fig:6}
\end{figure}

\begin{figure}[t]
\includegraphics[width=\columnwidth]{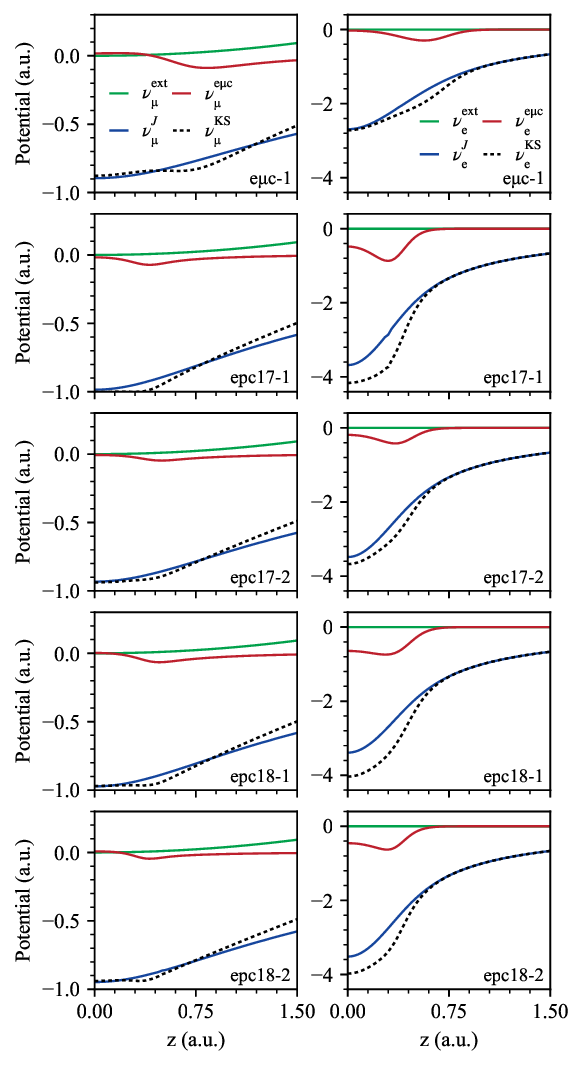}%
\caption{The functional-derived $\nu_\mathrm{e}^\mathrm{KS}$ and $\nu_\mathrm{\upmu}^\mathrm{KS}$, and their components, $\nu_\mathrm{e}^\mathrm{ext}$, $\nu_\mathrm{e}^J$, $\nu_\mathrm{e}^\mathrm{e\upmu c}$, and $\nu_\mathrm{\upmu}^\mathrm{ext}$, $\nu_\mathrm{\upmu}^J$, $\nu_\mathrm{\upmu}^\mathrm{e\upmu c}$, respectively, for Mu-DHT (see Eq. (\ref{eq:5}) for more details), deduced from the SCF-derived $\rho_\mathrm{e}$ and $\rho_\mathrm{\upmu}$. {The muonic and electronic potentials are shown in the left- and right-hand panels, respectively.} Since all the potentials are isotropic, the direction of the z-axis is arbitrary and the center of the coordinate system is placed at the joint center of the double harmonic traps.}
\label{fig:7}
\end{figure}

In order to apply the functionals to the model, we modified our in-house version of the NEO code \cite{webb_MulticonfigurationalNuclearelectronicOrbital_2002,goli_TwocomponentDensityFunctional_2022,pak_DensityFunctionalTheory_2007}, implemented in the GAMESS quantum computational package \cite{schmidt_GeneralAtomicMolecular_1993}, to include external harmonic potentials. The coupled KS equations were solved by expanding KS spatial orbitals in the previously described [7s7p7d/7s7p7d] basis set. For comparison purposes, the coupled KS equations were also solved without any correlation functional, abbreviated as the no-e(p/$\upmu$)c levels. Since there is no electron-electron correlation potential in the KS equations, these are practically equivalent to the MCHF/[7s7p7d/7s7p7d] computational level. The final numerical results are gathered in Table \ref{tab:1} and the resulting $\rho_\mathrm{e}$ and $\rho_\mathrm{p/\upmu}$ are depicted in Fig. \ref{fig:3} while the derived $\nu_\mathrm{e}^\mathrm{KS}$ and $\nu_\mathrm{p/\upmu}^\mathrm{KS}$ as well as their components are depicted in {Figs. \ref{fig:5}-\ref{fig:7}}.  

Let us first compare the functional-derived $\rho_\mathrm{e}$ and $\rho_\mathrm{p/\upmu}$ with the reference densities as one of the basic outputs of the SCF solutions of the coupled KS equations. Since the one-particle densities are isotropic, only a 1D plot of the densities along an arbitrary axis going through the joint centers of the double harmonic traps is given in Fig. \ref{fig:3}. In the case of H-DHT, the computed $\rho_\mathrm{e}$ with various correlation functionals, even that derived at the no-epc level, are almost superimposable on the reference $\rho_\mathrm{e}$. In contrast, the computed $\rho_\mathrm{p}$ is clearly overlocalized at the no-epc level but those derived using the correlation functionals of the epc series to a large extent reduce the overlocalization. Particularly, $\rho_\mathrm{p}$ derived from epc17-1 and epc18-1 are almost similar to the reference $\rho_\mathrm{p}$ in line with the fact that the numerical values of the parameters of these two functionals were optimized to reproduce the exact $\rho_\mathrm{p}$ in molecules \cite{yang_DevelopmentPracticalMulticomponent_2017,brorsen_AlternativeFormsTransferability_2018}. The worst result between the functionals is that of \emc-1 and this is also understandable since this functional has been primarily designed to cope with the overlocalization of $\rho_\mathrm{\upmu}$ in muonic molecules. In the case of Mu-DHT, the situation is to some extent different, and the computed $\rho_\mathrm{e}$ using various functional vary considerably. Oddly, the epc series is acting even worse than the {no-\emc} level when compared to the reference $\rho_\mathrm{e}$ and only \emc-1, to some extent, is able to reproduce the reference $\rho_\mathrm{e}$. The uncorrelated $\rho_\mathrm{\upmu}$ computed at the {no-\emc} level is evidently overlocalized and all functionals to some extent reduce the overlocalization although \emc-1 is clearly superior to the epc series. All these observations are promising and reveal the fact that the correlation functionals overcome the overlocalization of $\rho_\mathrm{p/\upmu}$, as they do in the case of real molecules. Also, we may claim that the model is sensitive enough to differentiate correctly between the functionals and reveal their special $m_\mathrm{PCP}$-dependent capabilities. 

Now, let us consider the KS energy contributions derived from each functional as given in Table \ref{tab:1}. The parameters of epc17-2 and epc18-2 were optimized to reproduce the zero-point energy of the proton in molecules thus one expects their performance in reproducing energetics to be superior in the epc series \cite{brorsen_AlternativeFormsTransferability_2018,brorsen_MulticomponentDensityFunctional_2017}. Indeed, both functionals reproduce the reference total energy of H-DHT and its components almost exactly. In contrast, epc17-1 and epc18-1 overestimate the absolute amount of the total energy as well as $T_\mathrm{e}^{s}$ and $J_\mathrm{e\textrm{-}p}$ as the two major components of the total energy. Interestingly, the performance of \emc-1 is much better than epc17-1 and epc18-1, and the only major error of this functional is in the correct reproduction of $T_\mathrm{p}^{s}$. In line with these observations, epc17-2, epc18-2 and \emc-1 are capable of recovering the reference $E_\mathrm{epc}$ quite precisely. Also, all functionals and even the no-epc level are capable of recovering the reference $\varepsilon_\mathrm{e}^\mathrm{KS}$, but concomitantly all of them severely underestimate the value of the reference $\varepsilon_\mathrm{p}^\mathrm{KS}$, $\sim 0.015$, predicting large negative values, $\sim -0.9$. As will be discussed subsequently in detail, the origin of this failure may be traced to the unsuccessful reproduction of the reference $\nu_\mathrm{p}^\mathrm{KS}$. Let us now turn to Mu-DHT, starting from the unexpected accurate recovery of the total reference energy by epc17-1 and epc18-1. Nevertheless, after an inspection of the energy components, it turns out that this must be due to error cancellations, since none reproduce the reference values of $T_\mathrm{e}^{s}$ and $J_\mathrm{e\textrm{-}\upmu}$ accurately. In contrast, epc17-2 and epc18-2 are not capable of recovering the total reference energy or the components properly. Expectedly, \emc-1 outperforms all the other functionals, however, even in this case the recovery of the reference $E_\mathrm{e\upmu c}$ is not very accurate. Finally, similar to the case of H-DHT, all functionals are capable of recovering the reference $\varepsilon_\mathrm{e}^\mathrm{KS}$ but none succeeds in reproducing the value of the reference $\varepsilon_{\upmu}^\mathrm{KS}$. 

To have a local view on the nature of correlations, {Fig. \ref{fig:5} depicts} the functional-derived correlation potentials deduced from the SCF-derived $\rho_\mathrm{e}$ and $\rho_\mathrm{p/\upmu}$ as well as the reference potentials. Since all potentials are isotropic, only 1D plots along an arbitrary axis going through the joint centers of the double harmonic traps are given in the {figure}. Figure \ref{fig:5} reveals that $\nu_\mathrm{p}^\mathrm{epc}$ and $\nu_{\upmu}^\mathrm{e\upmu c}$ deduced from the functionals of the epc series are quite similar and distinct from those deduced from \emc-1. The most prominent feature of this figure is the total failure of the functionals to reproduce the reference $\nu_\mathrm{p}^\mathrm{epc}$ and $\nu_{\upmu}^\mathrm{e \upmu c}$, and none are even remotely similar to the reference potentials. This is a disappointing result but let us stress that the situation is usually no better in the case of many seemingly successful electronic exchange-correlation functionals. This fact has been demonstrated long ago through a similar inversion process applied to the two-electron harmonium and helium atoms \cite{laufer_TestDensityfunctionalApproximations_1986,kais_DensityFunctionalsDimensional_1993,filippi_ComparisonExactApproximate_1994,umrigar_AccurateExchangecorrelationPotentials_1994,zhao_ElectronDensitiesKohnSham_1994,lam_VirialExchangeCorrelation_1998}. Accordingly, the reason(s) behind the successful reconstruction of $\rho_\mathrm{e}$ and $\rho_\mathrm{p/\upmu}$, and the KS energetics of the considered functionals is not tied to the successful reconstruction of the corresponding correlation potentials (for a relevant discussion, albeit for the electronic exchange-correlation functionals, see Refs. \cite{cruz_ExchangeCorrelationEnergy_1998,burke_UnambiguousExchangecorrelationEnergy_1998}). Most observations described above hold in the case of $\nu_\mathrm{e}^\mathrm{epc}$ and $\nu_\mathrm{e}^\mathrm{e\upmu c}$, depicted in {Fig. \ref{fig:5}}, with the major exception of clear similarities between the reference and functional-derived potentials particularly for \emc-1 derived potentials. This is a pleasant feature of the studied correlation functions, however, the role of $\nu_\mathrm{e}^\mathrm{e(p/\upmu)c}$ on shaping $\nu_\mathrm{e}^\mathrm{KS}$ is marginal and largely confined to the modification of the dominant $\nu_\mathrm{e}^{J}$ at small distances (\textit{vide infra}). Thus, future efforts to design proper electron-proton/muon functionals should mainly concentrate on the successful reproduction of $\nu_\mathrm{p}^\mathrm{epc}$ and $\nu_{\upmu}^\mathrm{e\upmu c}$. 

Let us now inspect the functional-derived effective potentials and their components, similar to those depicted in Fig. \ref{fig:4} for the reference KS system. Figure \ref{fig:6} depicts the functional-derived $\nu_\mathrm{e}^\mathrm{KS}$ and $\nu_\mathrm{p}^\mathrm{KS}$, and their components, respectively, computed from the SCF-derived $\rho_\mathrm{e}$ and $\rho_\mathrm{p}$ for H-DHT. Figure \ref{fig:7} depicts the same quantities for Mu-DHT revealing in conjunction with Fig. \ref{fig:6} that the derived $\nu_\mathrm{e}^\mathrm{KS}$ and $\nu_\mathrm{p/\upmu}^\mathrm{KS}$, and their components, are similar in the considered functionals particularly if only the epc series is taken into account. For all functionals, $\nu_\mathrm{e}^\mathrm{KS}$ has a minimum at the joint center of the harmonic traps and is almost superimposable on $\nu_\mathrm{e}^{J}$ with some marginal digressions induced by $\nu_\mathrm{e}^\mathrm{e(p/\upmu)c}$. Interestingly, the reference $\nu_\mathrm{e}^\mathrm{KS}$ depicted in Fig. \ref{fig:4} shares the same features with the functional-derived potentials, and this explains why the reference $\varepsilon_\mathrm{e}^\mathrm{KS}$ is properly recovered by the used functionals. In contrast, none of the functional-derived $\nu_\mathrm{p/\upmu}^\mathrm{KS}$ is even remotely similar to the reference potential which as discussed, is practically equal to $\nu_\mathrm{p/\upmu}^\mathrm{ext}$. This observation explains why the functional-derived $\varepsilon_\mathrm{p}^\mathrm{KS}$ values are severely disparate from those of the reference values, as disclosed previously. The dissimilarity stems from the fact that in contrast to the reference KS system, $\nu_\mathrm{p/\upmu}^\mathrm{e(p/\upmu)c}$ are unable to cancel the effect of $\nu_\mathrm{p/\upmu}^{J}$.         

We conclude that both epc series and \emc-1 are able to remedy the overlocalization of uncorrelated one-proton/muon densities in H/Mu-DHT, respectively. Moreover, they are also generally successful in reproducing the KS energetics and the electronic correlation potentials. Nonetheless, they are especially unable to reproduce the protonic/muonic KS orbital energies as well as the corresponding correlation potentials. Taking the fact that the model contains solely the electron-proton/muon correlation, the mentioned success confirms the capacity of these functionals to cope at least partly with this type of correlation. By the way, the mentioned failures also point to the fact that much room remains to improve the correlation functional design strategies. The model itself may serve as the target system for primary tests of any new electron-proton/muon correlation functional introduced in future studies.   

\begin{table*}[t]
  \centering
  \caption{The functional-derived $E_\mathrm{e(p/\upmu) c}$ computed using the uncorrelated, $E_\mathrm{e(p/\upmu) c}^\mathrm{no\textrm{-}e(p/\upmu)c}$, the SCF-derived, $E_\mathrm{e(p/\upmu) c}^\mathrm{SCF}$, (retrieved from Table \ref{tab:1}), and the reference, $E_\mathrm{e(p/\upmu) c}^\mathrm{ref}$, one-particle densities for H/Mu-DHT. The differences in the last two columns are given in kcal/mol.}
  \begin{ruledtabular}
    \begin{tabular}{c d{2.4}d{2.4}d{2.4} d{3.1}d{3.1}}
    \multicolumn{1}{c}{Functional} & \multicolumn{1}{c}{$E_\mathrm{e(p/\upmu) c}^\mathrm{no\textrm{-}e(p/\upmu)c}$} &\multicolumn{1}{c}{$E_\mathrm{e(p/\upmu) c}^\mathrm{SCF}$} & \multicolumn{1}{c}{$E_\mathrm{e(p/\upmu) c}^\mathrm{ref}$} & \multicolumn{1}{c} {$E_\mathrm{e(p/\upmu) c}^\mathrm{no\textrm{-}e(p/\upmu)c}-E_\mathrm{e(p/\upmu) c}^\mathrm{ref}$} & \multicolumn{1}{c} {$E_\mathrm{e(p/\upmu) c}^\mathrm{SCF}-E_\mathrm{e(p/\upmu) c}^\mathrm{ref}$} \\ \hline
          & \multicolumn{5}{c}{H} \\ \cmidrule{2-6}
    epc17-1 & -0.0344 & -0.0636 & -0.0565 & 13.9 & -4.5   \\
    epc17-2 & -0.0189 & -0.0314 & -0.0346 & 9.9  &  2.0   \\
    epc18-1 & -0.0208 & -0.0532 & -0.0427 & 13.8 & -6.6   \\
    epc18-2 & -0.0140 & -0.0297 & -0.0293 & 9.5  & -0.3   \\
    \emc-1  & -0.0048 & -0.0314 & -0.0191 & 9.0  & -7.7   \\
          & \multicolumn{5}{c}{Mu} \\ \cmidrule{2-6}
    epc17-1 & -0.0524 & -0.0696 & -0.0406 & -7.4 & -18.2   \\
    epc17-2 & -0.0379 & -0.0439 & -0.0367 & -0.7 & -4.5    \\
    epc18-1 & -0.0469 & -0.0637 & -0.0461 & -0.5 & -11.0   \\
    epc18-2 & -0.0312 & -0.0399 & -0.0258 & -3.4 & -8.8    \\
    \emc-1  & -0.0337 & -0.0619 & -0.0700 & 22.8 &  5.1    \\
    \end{tabular}%
    \end{ruledtabular}
  \label{tab:2}%
\end{table*}%

\begin{figure}[t]
\includegraphics[width=\columnwidth]{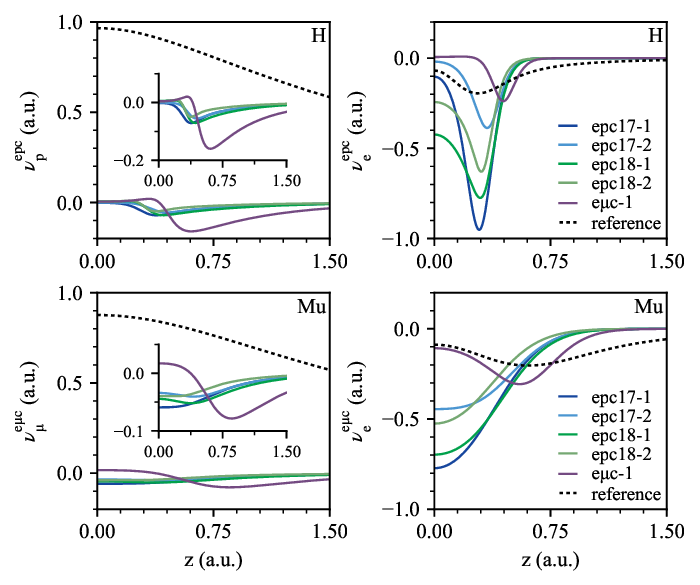}%
\caption{The reference and functional-derived {correlation potentials} for H/Mu-DHT. The reference correlation potentials are the exact results, depicted previously in Fig. \ref{fig:4}, while the others have been deduced using the considered electron-proton/muon correlation functionals. The potentials deduced for each functional are computed using {the reference/exact $\rho_\mathrm{p/\upmu}$ and} $\rho_\mathrm{e}$. {The insets are zoomed views of the functional-derived potentials.} Since all the potentials are isotropic, the direction of the z-axis is arbitrary and the center of the coordinate system is placed at the joint center of the double harmonic traps.}
\label{fig:8}
\end{figure}

\subsection{Disentangling the density-driven and the intrinsic errors of the functionals}

In the previous {subsection}, the quality of the correlation functionals was evaluated through the results gained from the SCF solution of the KS coupled equations. However, it is well-documented that this procedure is prone to two independent sources of errors \cite{kim_UnderstandingReducingErrors_2013,sim_QuantifyingDensityErrors_2018,vuckovic_DensityFunctionalAnalysis_2019}. One comes from the approximate nature of the functional itself, called intrinsic functional errors, and the other stems from the approximate one-particle densities derived from the SCF procedure, called density-driven errors. In this {subsection}, we try to disentangle these two errors to access the intrinsic quality of the studied correlation functionals by considering the associated potentials as well as the computed energies.            

{Figure \ref{fig:8} depicts} the functional-derived correlation potentials deduced from the reference $\rho_\mathrm{p/\upmu}$ and $\rho_\mathrm{e}$, respectively. In the case of H-DHT, through comparison with {Fig. \ref{fig:5}}, it becomes evident that for each functional, $\nu_\mathrm{p}^\mathrm{epc}$ and $\nu_\mathrm{e}^\mathrm{epc}$ deduced from reference and SCF-derived $\rho_\mathrm{p}$ and $\rho_\mathrm{e}$ are almost superimposable. The justification is that the reference and the SCF-derived $\rho_\mathrm{e}$ of all the studied functionals are almost superimposable, whereas the corresponding $\rho_\mathrm{p}$ are quite localized around the joint center of the traps (see Fig. \ref{fig:3}). Accordingly, $\rho_\mathrm{e}$ has a much more important role in shaping the topographies of $\nu_\mathrm{p}^\mathrm{epc}$ and $\nu_\mathrm{e}^\mathrm{epc}$, and $\rho_\mathrm{p}$ only affects the potentials around the joint center of the harmonic traps. The situation is completely different in the case of Mu-DHT as the epc series derived $\nu_\mathrm{\upmu}^\mathrm{e\upmu c} $ and $ \nu_\mathrm{e}^\mathrm{e\upmu c}$ vary considerably upon replacing the reference $\rho_\mathrm{\upmu}$ and $\rho_\mathrm{e}$ with {their SCF-derived counterparts}. The justification lies in the pronounced differences between the reference and SCF-derived $\rho_\mathrm{e}$ as well as the less localized nature of $\rho_\mathrm{\upmu}$ compared to $\rho_\mathrm{p}$ (see Fig. \ref{fig:3}). {Figure \ref{fig:8} also reveals} that the minima of the epc series derived $\nu_\mathrm{\upmu}^\mathrm{e\upmu c} $ and $ \nu_\mathrm{e}^\mathrm{e\upmu c}$ are all displaced to the joint center of the traps upon using the reference $\rho_\mathrm{\upmu}$ and $\rho_\mathrm{e}$ instead of the SCF-derived {one-particle densities}. Evidently, the lighter Mu-DHT is a more sensitive probe to check the density-dependence of the correlation potentials than H-DHT. Interestingly, in contrast to the $\nu_\mathrm{\upmu}^\mathrm{e\upmu c} $ and $ \nu_\mathrm{e}^\mathrm{e\upmu c}$ derived from the epc series of functionals, those of \emc-1 are less sensitive to the replacement of the SCF-derived $\rho_\mathrm{\upmu}$ and $\rho_\mathrm{e}$ with the reference densities. One may conclude at this stage that among the considered functionals, \emc-1 is least affected by density errors, as far as one is concerned with the correlation potentials.
 
Table \ref{tab:2} offers the functional-derived {$E_\mathrm{e(p/\upmu)c}$} for H/Mu-DHT computed using the uncorrelated, the SCF-derived (retrieved from Table \ref{tab:1}) and the reference one-particle densities. Inspection of the last two columns of the table demonstrates that in the case of H-DHT the computed $E_\mathrm{epc}$ from the epc series of functionals using the SCF-derived $\rho_\mathrm{p}$ and $\rho_\mathrm{e}$ is much more accurate than those derived using the uncorrelated densities. The same is true about the computed $E_\mathrm{e\upmu c}$ for Mu-DHT using \emc-1, however, surprisingly, the epc series of functionals seem to work better when using the uncorrelated $\rho_\mathrm{\upmu}$ and $\rho_\mathrm{e}$. This is probably the result of some type of error cancellation, though it is hard to pinpoint its exact nature. Nevertheless, these observations once again witness the model’s capacity to differentiate between the epc series and \emc-1 functionals, revealing their system-specific performance. With these results at hand, we may now proceed to compute the density-driven and intrinsic functional errors in total energies.  

In order to disentangle the density-driven and intrinsic functional errors of the functionals in reproducing the total energies, the following equation is used where for consistency, the notation is borrowed from the original literature \cite{kim_UnderstandingReducingErrors_2013,sim_QuantifyingDensityErrors_2018,vuckovic_DensityFunctionalAnalysis_2019}: 
\begin{align}\label{eq:7}
\Delta E =&\tilde{E}\left[ {{{\tilde{\rho }}}_\mathrm{e}},{{{\tilde{\rho }}}_\mathrm{p/\upmu }} \right]-E\left[ {{\rho }_\mathrm{e}},{{\rho }_\mathrm{p/\upmu }} \right] \nonumber \\ 
=&\left( \tilde{E}\left[ {{{\tilde{\rho }}}_\mathrm{e}},{{{\tilde{\rho }}}_\mathrm{p/\upmu }} \right]-\tilde{E}\left[ {{\rho }_\mathrm{e}},{{\rho }_\mathrm{p/\upmu }} \right] \right) \nonumber \\ 
&+\left( \tilde{E}\left[ {{\rho }_\mathrm{e}},{{\rho }_\mathrm{p/\upmu }} \right]-E\left[ {{\rho }_\mathrm{e}},{{\rho }_\mathrm{p/\upmu }} \right] \right) \nonumber \\ 
=&\Delta {{E}_\mathrm{D}}+\Delta {{E}_\mathrm{F}}.
\end{align}
In this equation, tildes are used to denote the SCF-derived total energies and the one-particle densities while those without tildes are the references; $\Delta {{E}_\mathrm{D}}$ is the density-driven error and $\Delta {{E}_\mathrm{F}}$ is the intrinsic error of a functional in reproducing the total exact energy. Table \ref{tab:3} offers $\Delta {{E}}$, $\Delta {{E}_\mathrm{D}}$ and $\Delta {{E}_\mathrm{F}}$ for the five considered functionals using the data from Tables \ref{tab:1} and \ref{tab:2}. In the case of H-DHT, $\Delta {{E}_\mathrm{F}}$ is the dominant error for epc17-1 and epc18-1 while for the remaining functionals and at the no-epc level, the values of $\Delta {{E}_\mathrm{D}}$ are comparable to those of $\Delta {{E}_\mathrm{F}}$. As discussed previously, epc17-2 and epc18-2 are capable of reproducing the KS energetics of H-DHT accurately and while their overall $\Delta {{E}}$ is small compared to the other considered functionals, $\Delta {{E}_\mathrm{D}}$ seems to have a non-negligible contribution to this success. In the case of Mu-DHT, $\Delta {{E}_\mathrm{D}}$ is also a major source of error for all the considered functionals except for \emc-1, which as discussed is capable of reproducing KS energetics accurately, although even for this functional its contribution is not negligible. Interestingly, in most cases, the signs of $\Delta {{E}_\mathrm{D}}$ and $\Delta {{E}_\mathrm{F}}$ are different and $\Delta {{E}_\mathrm{D}}$ masks the larger intrinsic errors of the considered functionals. We conclude that the success of the considered correlation functionals to reproduce KS energetics is not solely revealing their intrinsic performance, but partly it is the result of the density-driven errors.  

\begin{table}[t]
  \centering
  \caption{The density-driven, $\Delta {{E}_\mathrm{D}}$, and the intrinsic, $\Delta {{E}_\mathrm{F}}$, and the total energy, $\Delta {{E}}$, errors of the considered functionals computed for H/Mu-DHT (see Eq. (\ref{eq:7}) for the definitions). The numbers are given in kcal/mol.}
  \begin{ruledtabular}
    \begin{tabular}{c d{3.1}d{3.1}d{3.1}}
    \multicolumn{1}{c}{Method} & \multicolumn{1}{c}{$\Delta {{E}_\mathrm{D}}$} &\multicolumn{1}{c}{$\Delta {{E}_\mathrm{F}}$} & \multicolumn{1}{c}{$\Delta E$} \\ \hline
          & \multicolumn{3}{c}{H} \\ \cmidrule{2-4}
    no-epc  &   -7.1 & 20.7 & 13.7 \\   
    epc17-1 & -2.0  & -14.7 & -16.7 \\
    epc17-2 & -0.8  & -1.0  & -1.8  \\
    epc18-1 & -1.9  & -6.1  & -7.9  \\
    epc18-2 & -1.4  & 2.4   & 1.0   \\
    \emc-1  & -5.1  & 8.8   & 3.7   \\
          & \multicolumn{3}{c}{Mu} \\ \cmidrule{2-4}
    no-\emc &  -18.4  & 55.5  & 37.1 \\    
    epc17-1 & -31.3 & 30.0  & -1.3 \\
    epc17-2 & -21.0 & 32.5  & 11.5 \\
    epc18-1 & -24.0 & 26.6  & 2.6  \\
    epc18-2 & -24.5 & 39.4  & 14.9 \\
    \emc-1  & -4.0  & 11.6  & 7.6  \\
    \end{tabular}%
    \end{ruledtabular}
  \label{tab:3}%
\end{table}%

\section{Conclusions and prospects}

The electron-PCP correlation functional design is a vital part of MCDFT, however, in contrast to eDFT \cite{barth_BasicDensityFunctionalTheory_2004,cohen_InsightsCurrentLimitations_2008,cohen_ChallengesDensityFunctional_2012,burke_PerspectiveDensityFunctional_2012,becke_PerspectiveFiftyYears_2014,jones_DensityFunctionalTheory_2015,pribram-jones_DFTTheoryFull_2015,wasserman_ImportanceBeingInconsistent_2017,mardirossian_ThirtyYearsDensity_2017,verma_StatusChallengesDensity_2020,kaplan_PredictivePowerExact_2023}, based on the number of currently available functionals and the used design strategies, it is a much less developed research area. Particularly entertaining is the accessibility to few-body systems containing solely electron-PCP correlation to calibrate the performance of the newly proposed correlation functionals. The two-particle electron-PCP system within the double harmonic trap, proposed in the present study, is a first step in this direction. Notwithstanding, by adding more particles (electrons and/or PCPs) to the model, one may consider the interplay between various types of correlations, i.e., electron-electron, electron-PCP and PCP-PCP. Also, a more comprehensive study of the model itself in a larger domain of the parameters namely, PCP’s mass and the frequency of oscillation, may widen its application beyond protonic and muonic systems and the usual ambient conditions. Another interesting possibility is the application of such simple model systems to gain insight into the design of efficient electron-PCP correlation functionals. As the present study demonstrates, the currently available local electron-proton/muon functionals, with all their achievements, yet have clear weaknesses like the inability to yield correct protonic/muonic correlation potentials and corresponding orbital energies. Even more, part of their success in predicting correct energetics seems to be the result of the density-driven errors that mask the intrinsic shortcomings of the used functionals. All these point to the fact that much remains to be done in this research area.

% \section*{Supplementary material}
% The supplementary material contains ...

\begin{acknowledgments}
S.S. would like to acknowledge the generous access to the computational resources of SARMAD Cluster at Shahid Beheshti University.
\end{acknowledgments}

\appendix

\section{The basis set design for the H/Mu-DHT}\label{sec:a1}

The uncontracted [7s7p7d/7s7p7d] Gaussian basis set was designed as follows. The exponents of the Gaussian functions were derived through a non-linear energy optimization at the MCHF/[7s/7s] level and then used without any modification also for the p- and d-type Gaussian functions in the [7s7p7d/7s7p7d] basis set. Two series of exponents were derived, one for the proton’s mass and the other for the muon’s mass within the corresponding frequencies of oscillation, all given in Table \ref{tab:a1}. This basis set is also used without further modifications to solve the MCKS equations for the model. 

\begin{table}[t]
  \centering
  \caption{The exponents of the s-, p- and d-type Gaussian functions in the [7s7p7d/7s7p7d] basis set.}
  \begin{ruledtabular}
    \begin{tabular}{c d{2.4}d{2.4}}
    \multicolumn{1}{c}{Exponent} & \multicolumn{1}{c}{Electron} & \multicolumn{1}{c}{Proton/Muon} \\ \hline
          & \multicolumn{2}{c}{H} \\ \cmidrule{2-3}
    1     & 15.3638 & 22.5435 \\
    2     & 4.6647 & 13.6540 \\
    3     & 0.6693 & 6.9399 \\
    4     & 0.0573 & 5.7960 \\
    5     & 1.6563 & 4.1400 \\
    6     & 0.1286 & 2.0700 \\
    7     & 0.2893 & 1.0350 \\
          & \multicolumn{2}{c}{Mu} \\ \cmidrule{2-3}
    1     & 7.2212 & 15.8560 \\
    2     & 3.2589 & 10.4040 \\
    3     & 0.6472 & 7.0649 \\
    4     & 0.0574 & 5.2960 \\
    5     & 1.4523 & 4.1400 \\
    6     & 0.1287 & 2.0700 \\
    7     & 0.2879 & 1.0350 \\
    \end{tabular}%
    \end{ruledtabular}
  \label{tab:a1}%
\end{table}%

\section{The computational procedure of deducing the one-particle densities of HCN and MuCN}\label{sec:a2}

In the case of HCN and MuCN, since we do not have access to the exact ground state wavefunctions, the “reference” $\rho_\mathrm{p/\upmu}$ was derived by employing the double-adiabatic approximation \cite{goli_TwocomponentDensityFunctional_2022,bonfa_EfficientReliableStrategy_2015}. Within the context of this approximation, the proton/muon experiences an effective field produced by electrons and the clamped carbon and nitrogen nuclei. The resulting single-particle Schr\"{o}dinger equation is solved using the generalized Numerov method \cite{kuenzer_PushingLimitGridbased_2016}, and the derived one-proton/muon wavefunction was then squared to yield the reference $\rho_\mathrm{p/\upmu}$. To construct the effective potential, the electronic Schr\"{o}dinger equation was first solved at the B3LYP/pc-2 level \cite{becke_DensityFunctionalThermochemistry_1993a,jensen_PolarizationConsistentBasis_2001}, for the fixed equilibrium arrangement of the clamped carbon and nitrogen nuclei, while the clamped proton was placed at various points of a 3D cubic grid (for the grid details see Ref. \cite{goli_TwocomponentDensityFunctional_2022}). The uncorrelated $\rho_\mathrm{p/\upmu}$ was obtained at the B3LYP/pc-2//no-epc/14s14p14d and B3LYP/pc-2//no-\emc/14s14p14d levels, respectively. Note {that in this} notation no-e(p/$\upmu$)c implies that no electron-proton/muon correlation functional is used in the MCKS equations while [14s14p14d] is the protonic/muonic basis set used to expand the protonic/muonic KS spatial orbital (for the basis set details see Ref. \cite{goli_TwocomponentDensityFunctional_2022}). The reference $\rho_\mathrm{e}$ for XCN species were computed at the B3LYP/pc-2//epc17-1/14s14p14d and B3LYP/pc-2//\emc-1/14s14p14d levels wherein epc17-1 and \emc-1 are the electron-proton/muon correlation functionals \cite{yang_DevelopmentPracticalMulticomponent_2017,goli_TwocomponentDensityFunctional_2022}, while the uncorrelated $\rho_\mathrm{e}$ were derived at the B3LYP/pc-2//no-epc/14s14p14d and B3LYP/pc-2//no-\emc/14s14p14d levels. 

\section{The explicit forms of the considered functionals}\label{sec:a3}

The used functional forms are given below for comparison:
\begin{widetext}
\begin{equation*}
{{E}_\mathrm{epc17}}\left[ {{\rho }_\mathrm{e}},{{\rho }_\mathrm{p}} \right]=-\int{d\vec{r}\frac{{{\rho }_\mathrm{e}}\left( {{{\vec{r}}}_\mathrm{e}} \right){{\rho }_\mathrm{p}}\left( {{{\vec{r}}}_\mathrm{p}} \right)}{a-b{{\left( {{\rho }_\mathrm{e}}\left( {{{\vec{r}}}_\mathrm{e}} \right){{\rho }_\mathrm{p}}\left( {{{\vec{r}}}_\mathrm{p}} \right) \right)}^{{1}/{2}}}+c{{\rho }_\mathrm{e}}\left( {{{\vec{r}}}_\mathrm{e}} \right){{\rho }_\mathrm{p}}\left( {{{\vec{r}}}_\mathrm{p}} \right)}},
\end{equation*}
\begin{equation*}
{{E}_\mathrm{epc18}}\left[ {{\rho }_\mathrm{e}},{{\rho }_\mathrm{p}} \right]=-\int{d\vec{r}\frac{{{\rho }_\mathrm{e}}\left( {{{\vec{r}}}_\mathrm{e}} \right){{\rho }_\mathrm{p}}\left( {{{\vec{r}}}_\mathrm{p}} \right)}{{a}'-{b}'{{\left( {{\rho }_\mathrm{e}}{{\left( {{{\vec{r}}}_\mathrm{e}} \right)}^{{1}/{3}}}+{{\rho }_\mathrm{p}}{{\left( {{{\vec{r}}}_\mathrm{p}} \right)}^{{1}/{3}}} \right)}^{3}}+{c}'{{\left( {{\rho }_\mathrm{e}}{{\left( {{{\vec{r}}}_\mathrm{e}} \right)}^{{1}/{3}}}+{{\rho }_\mathrm{p}}{{\left( {{{\vec{r}}}_\mathrm{p}} \right)}^{{1}/{3}}} \right)}^{6}}}},
\end{equation*}
\begin{align}\label{eq:a3:1}
{{E}_\mathrm{e\upmu c\text{-}1}}\left[ \rho _\mathrm{e}^{\upalpha},\rho _\mathrm{e}^{\upbeta},{{\rho }_{\upmu}} \right]=-\int &d\vec{r} \frac{2\rho _\mathrm{e}^{\upalpha}\left( {{{\vec{r}}}_\mathrm{e}} \right){{\rho }_{\upmu}}\left( {{{\vec{r}}}_{\upmu}} \right)-\rho _\mathrm{e}^{\upalpha}\left( {{{\vec{r}}}_\mathrm{e}} \right){{\rho }_{\upmu}}{{\left( {{{\vec{r}}}_{\upmu}} \right)}^{{3}/{2}}}}{1+8\rho _\mathrm{e}^{\upalpha}\left( {{{\vec{r}}}_\mathrm{e}} \right){{\rho }_{\upmu}}\left( {{{\vec{r}}}_{\upmu}} \right)+4\rho _\mathrm{e}^{\upalpha}\left( {{{\vec{r}}}_\mathrm{e}} \right){{\rho }_{\upmu}}{{\left( {{{\vec{r}}}_{\upmu}} \right)}^{{3}/{2}}}}  \nonumber \\
&+\frac{2\rho _\mathrm{e}^{\upbeta}\left( {{{\vec{r}}}_\mathrm{e}} \right){{\rho }_{\upmu}}\left( {{{\vec{r}}}_{\upmu}} \right)-\rho _\mathrm{e}^{\upbeta}\left( {{{\vec{r}}}_\mathrm{e}} \right){{\rho }_{\upmu}}{{\left( {{{\vec{r}}}_{\upmu}} \right)}^{{3}/{2}}}}{1+8\rho _\mathrm{e}^{\upbeta}\left( {{{\vec{r}}}_\mathrm{e}} \right){{\rho }_{\upmu}}\left( {{{\vec{r}}}_{\upmu}} \right)+4\rho _\mathrm{e}^{\upbeta}\left( {{{\vec{r}}}_\mathrm{e}} \right){{\rho }_{\upmu}}{{\left( {{{\vec{r}}}_{\upmu}} \right)}^{{3}/{2}}}}.
\end{align}
% % \end{widetext}
In these expressions, $\int{d\vec{r}\equiv \int{d{{{\vec{r}}}_\mathrm{e}}}}\int{d{{{\vec{r}}}_\mathrm{p/\upmu }}} \; \delta ( {{{\vec{r}}}_\mathrm{e}}-{{{\vec{r}}}_\mathrm{p/\upmu }} )$, and ${\rho }_\mathrm{e}=\rho _\mathrm{e}^{\upalpha}+\rho _\mathrm{e}^{\upbeta}$ where $\rho_\mathrm{e}^{\upalpha}$ and $\rho_\mathrm{e}^{\upbeta}$ stand for the spin-up and spin-down electron densities, respectively. The constants in the epc series have been derived through regression procedures: $a=2.35$, $b=2.4$, $c=3.2$ for epc17-1 \cite{yang_DevelopmentPracticalMulticomponent_2017}, $a=2.35$, $b=2.4$, $c=6.6$ for epc17-2 \cite{brorsen_MulticomponentDensityFunctional_2017}, $a'=1.8$, $b'=0.1$, $c'=0.03$ for epc18-1 \cite{brorsen_AlternativeFormsTransferability_2018}, and $a'=3.9$, $b'=0.5$, $c'=0.06$ for epc18-2 \cite{brorsen_AlternativeFormsTransferability_2018}. Also, it is noteworthy that the kernel of \emc-1 functional for the closed-shell case, where $\rho_\mathrm{e}^{\upalpha}=\rho_\mathrm{e}^{\upbeta}=\frac{\rho_\mathrm{e}}{2}$, reduces to the following form  \cite{goli_TwocomponentDensityFunctional_2022}:
% \begin{widetext}
\begin{align}\label{eq:a3:2}
{{E}_\mathrm{e\upmu c\text{-}1}} \left[ \rho _\mathrm{e},{{\rho }_{\upmu }} \right] =-\int{d\vec{r}\frac{2\rho _\mathrm{e}\left( {{{\vec{r}}}_\mathrm{e}} \right){{\rho }_{\upmu }}\left( {{{\vec{r}}}_{\upmu }} \right)-\rho _\mathrm{e}\left( {{{\vec{r}}}_\mathrm{e}} \right){{\rho }_{\upmu }}{{\left( {{{\vec{r}}}_{\upmu }} \right)}^{3/2}}}{1+4\rho _\mathrm{e}\left( {{{\vec{r}}}_\mathrm{e}} \right){{\rho }_{\upmu }}\left( {{{\vec{r}}}_{\upmu }} \right)+2\rho _\mathrm{e}\left( {{{\vec{r}}}_\mathrm{e}} \right){{\rho }_{\upmu }}{{\left( {{{\vec{r}}}_{\upmu }} \right)}^{3/2}}}}.
\end{align}
\end{widetext}

% \section*{AUTHOR DECLARATIONS}
% \subsection*{Conflict of Interest}
% The authors have no conflicts to disclose.

% \section*{Data Availability}
% The data that support the findings of this study are available within the article and its supplementary material.

% Reference section
% \bibliography{references.bib}
% \bibliography{references_linked.bib}

\end{document}